\documentclass[english,aip,jcp,superscriptaddress,showpacs,preprint]{revtex4-1} 
\usepackage{amsmath}
\usepackage{amssymb}
\usepackage{graphicx}
\usepackage{makeidx}
\usepackage[caption=false]{subfig}
\usepackage[usenames,dvipsnames]{color}

\newcommand{\bra}[1]{\left \langle #1 \right \vert}
\newcommand{\ket}[1]{\left \vert #1 \right \rangle}

\DeclareMathOperator{\sech}{sech}

\begin{document}
\title{A Unified Stochastic Formulation of Dissipative Quantum Dynamics. II. Beyond Linear Response of
Spin Baths}

\author{Chang-Yu Hsieh}
\affiliation{Department of Chemistry, Massachusetts Institute of Technology,
77 Massachusetts Avenue, Cambridge, MA 02139}
\affiliation{Singapore University of Technology and Design, Engineering Product Development, 8 Somapah Road, Singapore 487372}
\author{Jianshu Cao}
\affiliation{Department of Chemistry, Massachusetts Institute of Technology,
77 Massachusetts Avenue, Cambridge, MA 02139}
\affiliation{Singapore-MIT Alliance for Research and Technology (SMART) Centre, Singapore 138602}

\begin{abstract}
We use the ``generalized hierarchical equation of motion" proposed in Paper I
to study decoherence in a system coupled to a spin bath.  
The present methodology allows a systematic incorporation of higher order anharmonic effects of the bath
in dynamical calculations.  We investigate the leading order corrections to the linear response approximations for spin bath models.  Two types of spin-based environments are considered: (1) a bath of spins discretized from a continuous spectral density and (2) a bath of physical spins such as nuclear or electron spins.  
The main difference resides with how the bath frequency and the 
system-bath coupling parameters are chosen to represent an environment.
When discretized from a continuous spectral density, the system-bath coupling typically scales as $\sim 1/\sqrt{N_B}$ where $N_B$ is the number of bath spins.  This scaling suppresses the non-Gaussian characteristics of the spin bath
and justify the linear response approximations in the thermodynamic limit. 
For the physical spin bath models, system-bath couplings are directly deduced from spin-spin interactions 
and do not necessarily obey the $1/\sqrt{N_B}$ scaling.  It is not always possible to justify the linear response
approximations in this case.  Furthermore, if the spin-spin Hamiltonian and/or the bath parameters are highly symmetrical, these additional constraints generate non-Markovian and persistent dynamics that
is beyond the linear response treatments.
\end{abstract}

\maketitle

\section{Introduction}\label{sec:intr}

Understanding the dissipative quantum dynamics of a system embedded in a complex environment is an important 
topic across various sub-disciplines of physics and chemistry.  Significant progress in the understanding of 
condensed phase dynamics has been achieved within the context of a few prototypical models\cite{Leggett:1987wk,breuer:book,weiss:book} such as Caldeira-Leggett model and spin-boson model. The 
environment is often modeled as a bosonic bath characterized with a spectral density, 
from which bath-induced decoherence can be deduced.
This prevalent adoption of bosonic baths is based on following considerations: 
(1) the simple mathematical tractability of a Gaussian bath model, 
(2) the linear response of an environment is often sufficient to account for quantum dissipations and
(3) the spectral density can be extracted from the classical molecular dynamics simulations

Despite the above-mentioned merits, there exists scenarios in which the ``bosonization'' of an environment 
is inadequate. For instance, the electron-transfer reaction in condensed phase is often approximated with the
spin boson model. The abstract model treats the generic quantum environment as a set of harmonic oscillators,
which corresponds to taking only linear response of solvent effects outside a solvation shell while imposing
a ``harmonic approximation" on the vibrations modes inside the shell.  The anharmonicity and higher-order nonlinear 
response can be substantial when the donor-acceptor complex is strongly coupled to some low-frequency vibrational 
modes or present in a nonpolar liquids.  To better understand the anharmonic effects of the environment, several 
groups including ours\cite{Kryvohuz:2005jf,Wu:2001dd} have studied correlation functions of anharmonic oscillators and a generalized spin boson model with a bath of independent Morse or quartic oscillators.  Similarly, a spin bath can be viewed as an extreme limit of anharmonic oscillators and provides
additional insights into condensed phase dynamics.

On the other hand, physical spin bath models, corresponding to localized electron and/or nuclear spins, have received increased attentions due to ongoing interests in developing various solid-state quantum technologies
\cite{hsieh_rpp2012,Kloeffel:2013eg,gao_natphoton2015} under the ultralow temperature regime when interactions with the phonons or vibrational modes 
are strongly suppressed.  For these spin-based environments, the spectral density is no longer a convenient 
characterization of the bath.  Instead, each bath spin is explicitly specified with parameters $\{\omega_k, 
g_k\}$, the intrinsic energy scale and the system-bath coupling coefficients, respectively.  

In this work, we investigate corrections to the standard linear response treatment of quantum dissipations induced by a spin bath\cite{Prokofiev:420342}. 
To quantitatively capture these higher order responses,
we utilize our recently proposed generalized hierarchical equation of motion (gHEOM) method\cite{hsieh_cao_2016} 
to incorporate higher order cumulants of an influence functional into an extended
HEOM framework\cite{Tang:2016gh} through a stochastic formulation\cite{Stockburger:2002em,,Shao:2004et,Lacroix:2005in}.  Even though it is possible to derive
the gHEOM directly through the path integral influence functional approach\cite{Tanimura:2006ga,makri99}, we emphasize that the stochastic approach provides an extension to additional methodology developments such as
hybrid deterministic-stochastic algorithms\cite{Moix:2013jb}.  This is a direction we are pursuing.
Due to the enormous numerical efforts needed to generate accurate long-time results, we restrict the numerical
illustrations in the short-time limit. Recently, our group\cite{Cerrillo:2014gl} and others have proposed methods 
to construct the memory kernel of a generic bath from numerically exact short-time dynamical results.
Hence, the gHEOM provides an invaluable tool to capture the anharmonicity and non-Gaussian effects 
of a generic quantum environment when 
used along with these other methods to correctly reproduce long-time results.  
Furthermore, starting from the stochastic formalism or the gHEOM, it also serves as a starting point to derive
master equations\cite{Jang:2002ds} incorporating these higher order nonlinear effects and can be more efficiently solved to extract long-time results.

As alluded earlier, we cover both scenarios:  
the spin bath as a specific realization of an anharmonic condensed-phase environment and
the physical spin bath.
In this work, we always explicitly take a spin bath as a collection of finite number of spins.  For physical
spin bath models, this restriction reflects the reality that there could only be a finite number of spins in the surrounding environment.  For an anharmonic condensed phase environment, if we simply take the spin bath as a realization of an infinitely large heat bath then it has been rigorously shown\cite{Suarez:1991be,makri99} that all higher order response function must vanish in the thermodynamic limit.  On the other hand, if we perform atomic simulation of solvents in a condensed phase, the anharmonicity can probably be attributed to a few prominent degrees of freedom.  Therefore, we restrict the number of bath spins in order to probe the effects of higher order response functions. Many earlier studies\cite{Caldeira:1993ud,Segal:2014ws,Wang:2012kk,Gelman:2004ji,Zhang:2015jl,Lu:2009hd} on the spin bath focused on the thermodynamic limit and restricted to analyze
the linear response only.  Some interesting phenomena include more coherent population dynamics\cite{Wang:2012kk}
in the nonadiabatic regime at elevated temperature and
the onset of negative thermal conductivity\cite{Segal:2014ws} 
in a molecular junction coupled to two large spin baths held at
different temperatures.  In App.~\ref{app:linear-resp}, we report our own investigation on differences between a spin bath and a bosonic bath in the linear response limit when the spin bath can be accurately mapped
onto an effective bosonic bath.

In this work, the main difference between the two types of environment 
comes down to how the parameters $\{\omega_k, g_k\}$
are assigned to each bath spin as explained later. In general, we find the anharmonicity is more pronounced at the low-frequency end when the spectral density for condensed phase environment
is the commonly used Ohmic form.  Therefore, a slow spin bath at low temperature 
could potentially pose the most difficulty for a linear response treatment of the bath.  
On the other hand, for a physical spin bath model, highly non-Markovian and persistent dynamics\cite{Bortz:2007ku,Wang:2013jx,Chen:2007kra,Seifert:2016vx,Breuer:2004vi} emerge under a combination of narrowly distributed bath parameters and highly symmetrical system-bath Hamiltonians. 
To accurately reproduce these results might require taking higher order response of the spin bath into account.

The paper is organized as follows. In Sec.\ref{sec:anharmonicity}, we introduce the spin bath models of interest.
In Sec.\ref{sec:method}, we provide a brief account of the stochastic formalism\cite{hsieh_cao_2016}
and how to use it to derive the gHEOM with a systematic inclusion of higher order cumulants of an influence functional.  
In Sec.\ref{sec:Res}, we first discuss an exactly solvable dephasing model to stress
the importance of higher order cumulant corrections and
present a benchmark to validate our numerical method then move on to study both
finite size representation of the condensed-phase environment and an Ising spin bath.  
A brief summary is given in Sec.\ref{sec:Con}. In App.\ref{app:stoch}, we provide additional materials on the stochastic derivation of the generalized HEOM.   
In App.\ref{app:linear-resp},  we investigate the linear response effects of the spin bath and physical signatures that one can use to distinguish a spin based
condensed-phase environment from a bosonic one.

\section{Spin Bath Models}\label{sec:anharmonicity}
We consider the following Hamiltonian in this work,
\begin{eqnarray}\label{eq:sysbath}
\hat H & = & \hat H_s + \hat H_B + \hat H_{\text{int}}\nonumber \\
& = & \frac{\epsilon}{2} \hat \sigma^z_0 + \frac{\Delta}{2} \hat \sigma^x_0 + \hat{H}_B + \hat{H}_{\text{int}},
\end{eqnarray} 
where $\hat H_B = \sum_{k>0} (\omega_k/2) \hat\sigma^z_k$ and the standard partition of the system (subscript s), 
bath (subscript B) and the mutual interaction (subscript int) is implied. The general spin-spin interaction takes
the form $\hat{\sigma}^\alpha_0 \hat{\sigma}^\beta_1$, where $\alpha / \beta$ denotes the Cartesian components of Pauli matrices.  Among the choices, most common system-bath interactions read,
\begin{eqnarray}\label{eq:interactionH}
\hat H_{\text{int}} = \left\{ 
\begin{array}{l}
\sum_{k>0} g_k \hat\sigma^z_0 \hat\sigma^x_k, \\
\sum_{k>0} g_k \hat\sigma^z_0 \hat\sigma^z_k, \\
\sum_{k>0} g_k (\hat\sigma^+_0 \hat\sigma^-_k+\hat\sigma^-_0 \hat\sigma^+_k), \\
\sum_{k>0} g_k (\hat\sigma^x_0 \hat\sigma^x_k+ \hat\sigma^y_0 \hat\sigma^y_k+ \hat\sigma^z_0 \hat\sigma^z_k).
\end{array}
\right.
\end{eqnarray}
In this work, we should explicitly consider first two interaction forms.
The first form is appropriate for modelling condensed-phase environment, while the second form is useful in
the decoherence study in quantum computing and related contexts.

\subsection{Anharmonic condensed-phase environment}\label{sec:morse}

One simple-model approach to investigate anharmonicity of a condensed-phase environment is to 
generalize the typical bosonic bath by substituting harmonic oscillators
with the quartic oscillators or Morse oscillators\cite{LopezLopez:2011en}.  
We briefly illustrate the steps to obtain a spin bath model corresponding to the low-energy spaces of 
a bath of Morse oscillators\cite{Kryvohuz:2005jf}.

We still begin with Eq.~(\ref{eq:sysbath}) but having a different bath part,
\begin{eqnarray}\label{eq:morse_H}
\hat{H}_B  & = & \sum_{k>0} \left( \frac{P_k^2}{2} + D_k \left( 1- e^{-\alpha_k X_k} \right)^2 \right), \nonumber \\
\hat{H}_{\text{int}} & = & \hat \sigma^z_0 \sum_{k>0} c_k X_k.
\end{eqnarray}
The $n$-th eigen-energy of the $k$-th Morse oscillator is given by
\begin{eqnarray}\label{eq:morse_en}
E_{n,k} = \omega_k \left( n + \frac{1}{2} \right) - \chi_k  \left( n + \frac{1}{2} \right)^2,
\end{eqnarray}
where the fundamental frequency $\omega_k = \alpha_k \sqrt{2D_k}$
and the anharmonicity factor $\chi_k =\alpha^2_k/2$.  Following Ref.~\onlinecite{LopezLopez:2011en},
one can characterize the anharmonicity of each mode by imposing a parameter, $\Lambda$ (the number of bound states
in a Morse oscillator).   Under the condition of fixed $\Lambda$, one gets
\begin{eqnarray}
D_k = \frac{\omega_k \Lambda}{2},  \,\, \alpha_k = \sqrt{\frac{\omega_k}{\Lambda}}, \,\,
\text{and} \,\, \chi_k = \frac{\omega_k}{2\Lambda},
\end{eqnarray}
for each Morse oscillator with a free parameter $\omega_k$.
As clearly implied in Eqs.~(\ref{eq:morse_H}) and (\ref{eq:morse_en}), 
the Morse potential and the energy level spacing 
smoothly reduce back to those of a harmonic oscillator in the limit of $\Lambda \rightarrow \infty$.
The recovered harmonic bath is characterized by
\begin{eqnarray}\label{eq:sb_bath}
\hat{H}_B  & = & \sum_{k>0} \left(\frac{P_k^2}{2} + \frac{1}{2}\omega_k^2 X_k^2 \right), \nonumber \\
\hat{H}_{\text{int}} & = & \hat \sigma^z_0 \sum_{k>0} c_k X_k.
\end{eqnarray}
On the other hand, by setting $\Lambda = 2$, an effective spin bath emerges.  
The Equation (\ref{eq:morse_H}) can now be cast as
\begin{eqnarray}\label{eq:ss_bath}
\hat{H}_B  & = & \sum_{k>0} \frac{\omega_k}{2} \hat{\sigma}^z_k, \nonumber \\
\hat{H}_{\text{int}} & = & \hat \sigma^z_0 \sum_{k>0} \frac{c_k}{\sqrt{2\omega_k}} \hat\sigma_k^x.,
\end{eqnarray}
which correspond to the first interaction form in Eq.~(\ref{eq:interactionH}).

The mapping of a generic anharmonic environment 
onto a spin bath is more universal than the specific example of Morse oscillators. 
A general approach to achieve the mapping is to formulate an 
influence functional of the bath then perform a cumulant expansion, which characterizes 
the bath-induced decoherence through multi-time correlation functions.  One then has a clear set of criteria
to construct a spin bath to reproduce the bath's response up to a specific cumulant expansion order.
This is achievable as a set of spins (or qubits) constitute a versatile quantum simulator\cite{Lloyd_sci96} 
that can simulate
other simple quantum systems.

\subsection{Physical Spin Bath}\label{sec:physpin}
In the present context, the spin bath is not just a conceptual model but represent the actual
spin-based environment composed of nuclear / electron spins in the surrounding medium of a physical system.
Depending on details regarding a system, spin-spin interactions could take on a number of different forms 
such as the Ising, flip-flop (or XX) and Heisenberg interactions in Eq.~(\ref{eq:interactionH}).  
For simplicity, we investigate the Ising interaction\cite{Krovi:2007iu,torron_kosloff_njp16} 
in addition to the first form of interaction in Eq.~(\ref{eq:interactionH}).

The physical spin bath must contain a finite number of bath spins.  In certain systems, such as electrically-gated 
quantum dots\cite{hsieh_rpp2012} in GaAs, there could be as many as $10^5 - 10^6$ nuclear spins within the quantum-dot confining
potential. While only a small fraction of bath spins are strongly coupled to the system; it is often possible
to make semi-classical approximations to simplify the calculations. 
On the other hand, for NV centers\cite{gao_natphoton2015} and related system, the relevant
spin bath contains only $10^1 - 10^2$ spins.  The bath could be potentially too small for semi-classical approximation and too large for a full dynamical simulations.  Although we have seen impressive advances in 
simulation methods\cite{Wang:2012kk,Dobrovitski:602306,AlHassanieh:2006kr,stanek:2013,Gelman:2004ji}
for large spin systems in the last decade, there still exists rooms for improvement.

\subsection{Spin Bath Parameters}\label{sec:2}

A finite-size bath model is fully characterized by pairs of parameters, $\{\omega_k, g_k\}$.
In modelling physical spin system, these parameters are often randomly drawn from narrow probability distributions as done and justified in earlier works\cite{shenvi_desousa_prb2005,loss_2003,zhang_dobrovitski_prb2006}. In particular, we will sample both $\omega_k$ and $g_k$ from the uniform distributions.
 
When addressing spin bath as a representation for anharmonic condensed phase environment, we consider an alternative assignment of parameters, 
$\{\omega_k, g_k\}$, by discretizing an Ohmic spectral density, $J(\omega)\propto \omega \exp(-\omega/\omega_c)$ according to the scheme given in Ref.~\onlinecite{makri99}.
In the thermodynamic limit, it has already been shown\cite{makri99} 
that the spin bath can be exactly mapped onto a bosonic one with a temperature-dependent
spectral density 
\begin{eqnarray}\label{eq:effspec}
J_{\text{eff}}(\omega) = \tanh\left(\frac{\beta\omega}{2}\right) J(\omega).
\end{eqnarray}
Our present focus is to investigate the nonlinear effects beyond the effective spectral density prescription.

\section{Methodology}\label{sec:method}

\subsection{Stochastic decoupling of many-body quantum dynamics}

We now present an approach to systematically incorporate the non-linear bath effects into the HEOM framework through
a stochastic calculus based derivation. 
Given the model described earlier, the exact quantum dynamics of the composite system (central spin and the bath) 
can be cast into a set of coupled stochastic differential equations,
\begin{eqnarray}
\label{eq:ito}
& & d\tilde\rho_s = -i dt \left[\hat{H}_s, \tilde\rho_s\right] 
- i dt \mathcal{B} (t) \left[ A, \tilde\rho_s \right]  \\
& & \, \, -\frac{i}{\sqrt 2}dW^{*} A \tilde\rho_s + \frac{i}{\sqrt 2} dV^{*} \tilde\rho_s A , \nonumber \\
\label{eq:ito2}
& & d\tilde\rho_B = -i dt \left[\hat{H}_B, \tilde\rho_B\right] 
+\frac{1}{\sqrt 2} dW \left( B - \mathcal{B}(t) \right) \tilde \rho_B \nonumber \\
& & \, \, + \frac{1}{\sqrt 2} dV \tilde \rho_B  \left( B- \mathcal B (t) \right).
\end{eqnarray}
where $\tilde\rho_{s/B}(t)$ refers to the stochastically evolved density matrices in the presence of the  white noises, implied by the
Wiener differential increments
$dW$ and $dV$, and
the bath-induced stochastic field acting on the system,
\begin{eqnarray}\label{eq:bfield}
\mathcal B (t) = 
\sum_k g_k \text{Tr}_B \left\{\tilde\rho_B(t) \sigma^x_k \right\}.
\end{eqnarray}
The reduced quantum dynamics of the central spin is recovered after averaging $\tilde\rho_s(t)$ 
over different noise realizations in the above equations, 
\begin{equation}\label{eq:formalavg}
\rho_s(t) = \mathcal{E}(\tilde\rho_s(t)).
\end{equation}
As implied in Eq.~(\ref{eq:ito}), the dissipative effects of the bath are completely captured by the interplay of the stochastic field $\mathcal{B}(t)$
and the white noises. To determined the stochastic field, one can use Eq.~(\ref{eq:ito2}) to derive a closed form expression for $\mathcal{B}(t)$ in the case of bosonic bath,
\begin{eqnarray}\label{eq:bosonfield}
\mathcal{B}(t) = \frac{1}{\sqrt{2}} \left( \int^t_0 dW(s)  \alpha(s)  + \int^t_0 dV(s) \alpha^*(s) \right),
\end{eqnarray}
where $\alpha(t) = \int^\infty_0 d\omega J(\omega) \left(\coth(\beta\omega/2) \cos(\omega t) - i \sin(\omega t)\right)$ is the two-time
correlation functions. 
For non-Gaussian bath models, such
as the spin bath, 
the stochastic field is determined by multi-time correlation functions as follows,
\begin{eqnarray}\label{eq:generalfield}
\mathcal B (t) 
& = & \frac{1}{\sqrt{2}}\int^t_0 dW(s) \Phi_{2,1}(t,s) + 
\frac{1}{\sqrt{2}}\int^t_0 dV(s) \Phi_{2,2}(t,2)  +  \nonumber \\
& & \left(\frac{1}{\sqrt 2}\right)^3 \int^t_0 \int^{s_1}_0 \int^{s_2}_0 dW(s_1) dW(s_2) dW(s_3) \Phi_{4,1}(t,s1,s2,s3) + \dots 
+\nonumber \\
& & \left(\frac{1}{\sqrt 2}\right)^3 \int^t_0 \int^{s_1}_0 \int^{s_2}_0 dV(s_1) dV(s_2) dV(s_3) \Phi_{4,8}(t,s1,s2,s3) + \dots ,
\end{eqnarray} 
where the definition on correlation functions $\Phi_{n,m}(t,t_1,\dots, t_{n-1})$  are delegated to the App.~\ref{app:stoch}. The equation (\ref{eq:generalfield}) assumes the odd-time correlation functions vanish with
respect to the initial thermal equilibrium state.
For now, we simply note that there are $2^{n-1}$ possible $n$-time correlation functions.  The second subscript $m=1,\dots 2^{n-1}$ labels these functions.  Not all of the $n$-time correlation functions are independent; every
correlation function and its complex conjugate version are counted separately in this case.  The motivation to distinguish the correlation functions is actually to differentiate all possible sequence of multiple integrations of noise variables associated with the $n$-time correlation functions as
shown in Eq.~(\ref{eq:generalfield}).  
In this equation, we have explicitly
expanded this formal expression up to the fourth-order correlation functions.

Substituting Eq.~(\ref{eq:generalfield}) into Eq.~(\ref{eq:ito}), one then derives a closed stochastic differential
equation for the central spin. Direct stochastic simulation schemes so far have only been proposed for the Gaussian baths when the stochastic field is exactly defined by the second cumulant as in Eq.~(\ref{eq:bosonfield}). In this case, the stochastic field can be combined with the white noises to define color noises with statistical properties specified by the bath's two-time correlation function.  When higher order cumulant terms are needed to properly characterize $\mathcal{B}(t)$, a direct stochastic
simulation becomes significantly more complicated.  In this study, we choose to convert Eqs.~(\ref{eq:ito}) and (\ref{eq:generalfield}) to a hierarchy of deterministic equations involving auxiliary density matrices.

\subsection{Generalized hierarchical equations of motion}\label{sec:gheom}

To derive the hierarchical equation, we begin with Eq.~(\ref{eq:ito}) and take a formal ensemble average of the noises to get
\begin{eqnarray}\label{eq:first-heom}
& & \frac{d\rho_s}{dt} = -i \left[\hat{H}_s, \rho_s\right] 
- i \left[ A, \mathcal{E}\left( \mathcal{B} (t)\tilde\rho_s \right) \right].
\end{eqnarray}
To arrive at the equation above, we invoke the relation of Eq.~(\ref{eq:formalavg}) and the fact $\mathcal{E}(dW) =
\mathcal{E}(dV) = 0$.  This deterministic equation now involves an auxiliary density matrix $\mathcal{E}(\mathcal{B}(t)\tilde\rho_s(t))$ that needs to be solved too.  Working out the equation of motion for the auxiliary density matrix (ADM), one is then required to define additional ADMs and a hierarchy forms.

Following a recently proposed scheme, we introduce a complete 
set of orthonormal functions $\{\phi_j(t)\}$ and express all
the multi-time correlation functions as
\begin{eqnarray}\label{eq:decomp-multitime}
\Phi_{n+1,m}(t,t_1,\dots,t_n) = \sum_{\pmb j} \chi^{n+1,m}_{\pmb j} \phi_{j_1}(t-t_1) \cdots \phi_{j_n}(t_n-t_1),
\end{eqnarray}
where $\pmb j = (j_1, \dots j_n)$.  Due to the completeness, one can also cast the derivatives of
the basis functions in the form,
\begin{eqnarray}
\frac{d}{dt} \phi_j(t) = \sum_{j'} \eta_{jj'} \phi_{j'}(t).  
\end{eqnarray}
Next we define the cumulant block matrices 
\begin{eqnarray}
\mathbf{A}_n = \left[ \begin{array}{l}
a^n_{1 \mathbf{j}_1} \cdots a^n_{1 \mathbf{j}_k}, 0, \dots \\
\vdots \\
a^n_{2^{n} \mathbf{k}_1} \cdots  a^n_{2^{n} \mathbf{j}_k}, 0, \dots
\end{array}\right],
\end{eqnarray}
where each composed of $2^{n+1}$ row vectors with indefinite size and the $0$ in each row implies all zeros beyond this point.
For instance, $A_1$ has two row vectors while $A_2$ has four row vectors etc. The $m$-th row vector of matrix $A_n$ contains matrix elements denoted by $(a^n_{m \mathbf{j}_1}, a^n_{m \mathbf{j}_2} , \dots a^n_{m \mathbf{j}_k})$.  Each of this matrix element can be
further interpreted by
\begin{eqnarray}\label{eq:adef}
a^n_{m \mathbf{j}} \equiv \left(\frac{1}{\sqrt{2}}\right)^{n} \int^t_0 \int^{s_1}_0 \dots \int^{s_{n-1}}_0 dU(s_1) \dots dU(s_{n}) \phi_{j_1}(t-s_1) 
\phi_{j_2}(s_2-s_1) \cdots \phi_{j_{n}}(s_n-s_1), \nonumber \\
\end{eqnarray}
where $dU(s_j)$ can be either a $dW(s_j)$ or $dV(s_j)$ stochastic variable depending on index $m$.
With these new notations, the multi-time correlation functions in Eq.~(\ref{eq:generalfield}) are now concisely encoded by
\begin{eqnarray}
\mathcal{B}(t) = \sum_{n,m, \pmb j} \chi^{n+1,m}_{\pmb j} a^n_{m \pmb j}.
\end{eqnarray}
Now we introduce a set of ADM's  
\begin{eqnarray}
& & \rho^{\left[\mathbf A_1 \right] \left[ \mathbf A_2 \right] \left[ \mathbf A_3 \right]} \dots \equiv  \mathcal{E}\left( \prod_{n,m,k} a^n_{m \mathbf{j}_k} \tilde\rho_s(t)\right),
\end{eqnarray}
which implies the noise average over a product of all non-zero elements of each matrix $\mathbf{A}_i$ with the
stochastically evolved reduced density matrix of the central spin.  The desired reduced density matrix would correspond to the ADM in which all matrices are empty.  Furthermore, the very first ADM we discuss in Eq.~(\ref{eq:first-heom}) can be cast as
\begin{eqnarray}
\mathcal{E}\left( \mathcal{B} (t)\tilde\rho_s \right) = \sum_{n,m,\pmb j} \chi^{n+1,m}_{\pmb j}
\rho^{ \dots \left[\mathbf{A}_n\right] \dots }(t),
\end{eqnarray}
where each ADM, $\rho^{ \dots \left[\mathbf{A}_n\right] \dots }$, on the RHS of the equation carries only one non-trivial matrix element $a^{n}_{m,\pmb j}$ in
$\mathbf{A}_n$.
Finally, The hierarchical equations of motion for all ADMs can now be put in the following form,
\begin{eqnarray}\label{eq:gheom}
& & \partial_t \rho^{\left[ \mathbf A_1 \right] \left[ \mathbf A_2 \right] \left[ \mathbf A_3 \right]}  =  
-i \left[ H_s, \rho^{\left[ \mathbf A_1 \right] \left[ \mathbf A_2 \right] \left[ \mathbf A_3 \right]} \right] 
-i \sum_{n,m,\pmb j} \chi^{n,m}_{\pmb j} 
\left[ A, \rho^{\cdots \left[ \mathbf{A}_n + (m,\pmb j) \right] \cdots}  \right] \nonumber \\
& &  -i \sum_{n,m,\pmb j} 
\phi_{j_1}(0) A \rho^{\cdots \left[ \mathbf{A}_{n-1} + (m',\pmb{j}_1) \right]\left[ \mathbf{A}_n - (m,\pmb j) \right] \cdots} 
 -i \sum_{n,m,\pmb j} 
\phi_{j_1}(0) \rho^{\cdots \left[ \mathbf{A}_{n-1} + (m',\pmb{j}_1) \right]\left[ \mathbf{A}_n - (m,\pmb j) \right] \cdots} A \nonumber \\
& & + \sum_{n,m,\mathbf{j}} \eta_{\mathbf{j} \mathbf{j'}}  
\rho^{\cdots \left[ a^{n}_{m \mathbf{j}} \rightarrow a^{n}_{m \mathbf{j'}} \right] \cdots}
\end{eqnarray}
In this equation, we introduce a few compact notations that we now explain.  We use $\left[ \mathbf{A}_n \pm (m,\pmb j) \right]$ to mean adding or removing
an element $a^{n}_{m \pmb j}$ to the $m$-th row.  We also use  $\left[ a^{n}_{m \mathbf{j}} \rightarrow a^{n}_{m \mathbf{j'}} \right]$ to denote
a replacement of an element in the $m$-th row of $\mathbf{A}_n$. On the second line, we specify an element in a lower matrix given by $(m',\mathbf{j}_1)$.
The variable $\mathbf{j}_1$ implies removing the first element of the $\mathbf{j}$ 
array and the associated index $m'$ is determined by removing the first stochastic integral in Eq.~(\ref{eq:adef}). 
After the first term on the RHS of Eq.~(\ref{eq:gheom}), we only explicitly show the matrices $\mathbf{A}_n$ affected in each term of the equation.

\section{Results and Discussions}\label{sec:Res}
In the following numerical examples, we investigate the dynamics of a central spin coupled to a spin bath, Eq.~(\ref{eq:sysbath}).  First two forms of system-bath interactions in Eq.~(\ref{eq:interactionH}) 
will be addressed.  We use the gHEOM to simulate dynamics and
adopt the Chebyshev polynomials as the functional basis to interpolate 
high-dimensional multi-time correlation functions.
We note that there exists simpler choices\cite{Tang:2016gh} of functional basis when only the two-time correlation functions are needed as in the bosonic bath models.  We consider the standard initial conditions,
\begin{eqnarray}\label{eq:genrhoi}
\hat \rho(0) & = & \hat \rho_s(0) \otimes \hat \rho_B^{\text{eq}},
\end{eqnarray}
where the bath density matrix is simply a tensor product of the thermal equilibrium state for each individual mode.

Through the examples in this section, 
we investigate whether it is generally a valid idea to map a spin bath model
onto an effective bosonic one, such as obtained through a second order cumulant expansion of the spin bath's influence functional.  The gHEOM presented in Sec.~\ref{sec:gheom} will be used to quantify the contributions
of higher order cumulant corrections to the quantum dynamics of the central spin.

\subsection{Pure dephasing}\label{sec:dephasing}

We first analyze a pure dephasing model\cite{rao-pra2011}, i.e. $\Delta = 0$ in Eq.~(\ref{eq:sysbath}) and adopt the first interaction 
$\hat{H}_{\text{int}}=\hat\sigma^z_0 \sum_k g_k \hat\sigma_k^x$ in Eq.~(\ref{eq:interactionH}). This case can be analytically solved and 
provides insights into the higher order response functions of the bath.  
The coherence of the central spin can be expressed as
\begin{eqnarray}\label{eq:off-diagonal}
\bra{\uparrow}\rho_s(t)\ket{\downarrow} = \bra{\uparrow}\rho_s(0)\ket{\downarrow}e^{-i\epsilon t} e^{\Gamma(t)},
\end{eqnarray}
where the decoherence factor $\Gamma(t)$ 
reads, 
\begin{eqnarray}\label{eq:ss_coherence}
\Gamma(t) & = & \sum_k \ln \left \langle e^{iH^{(k)}_{+}t}e^{-i H^{(k)}_{-}t} \right \rangle \nonumber \\
& = & \sum_k \ln \left[1 - \frac{4 g_k^2}{\Omega_k^2}\left(1-\cos \Omega_k t \right) \right], \nonumber \\
& \approx & \sum_k \left[ \frac{4g_k^2}{\omega_k^2}\left( 1- \cos(\omega_k t)\right) - 
\left(\frac{4g_k^2}{\omega_k^2}\right)^2 \sin(\omega_k t) \left(\sin(\omega_k t) - \omega_k t\right) \right],
\end{eqnarray}
with
$\hat{H}^{(k)}_{\pm} = (\omega/2) \hat \sigma^z_k \pm g_k \hat \sigma^x_k $ and $\Omega_k = \omega_k\sqrt{1 + 
(4g_k^2/\omega_k^2)}$.  On the last line, 
we expand $\Gamma(t)$ to get the
two leading contributions with respect to $\lambda_k = 4g^2_k/\omega_k^2$.  These two terms correspond to the second order and fourth order
cumulant expansion of the influence functional for this particular model, respectively. 

Even with this simple case, one can draw important remarks regarding spin bath mediated decoherence.  First of all, the exact result in Eq.~(\ref{eq:ss_coherence}) implies the perturbations coming from a specific 
spin bath mode are modulated with an interaction renormalized frequency $\Omega_k$ as opposed to the bare frequency $\omega_k$, which is the way bosonic modes perturb a system through a linear coupling.
The origin of this interaction dressed $\Omega_k$ can be understood by inspecting the time evolution of Pauli matrices associated with individual spin modes,
\begin{eqnarray}\label{eq:ss_modes}
\hat{\sigma}^{\pm}_k(t) & = & e^{\pm i\omega t} \hat{\sigma}^{\pm}_k(0) \mp i g_k
 \int^t_0 d\tau e^{\pm i\omega_k(t-\tau)} \hat{\sigma}^{z}_k(\tau),\nonumber \\ 
\label{eq:ss_modes2}
\hat{\sigma}^{z}_k(t) &=& \hat{\sigma}^{z}_k(0) - i 2 g_k \int^t_0 d\tau \left(\hat{\sigma}^+_k(\tau) - \hat{\sigma}^-_k(\tau) \right).
\end{eqnarray}
By converting the raising and lowering Pauli matrices into the $x$ and $y$ Pauli matrices, it is 
obvious that different components of the bath spin get coupled together via the system-bath interaction in a non-trivial way and renormalize the frequency at which a spin precesses. 
There is no such coupling of the internal structure for harmonic oscillators due to the fundamental 
differences in the commutation properties of bosonic creation/annihilation operators and Pauli matrices for spins.

\begin{figure}
\centering
\includegraphics[width=\linewidth]{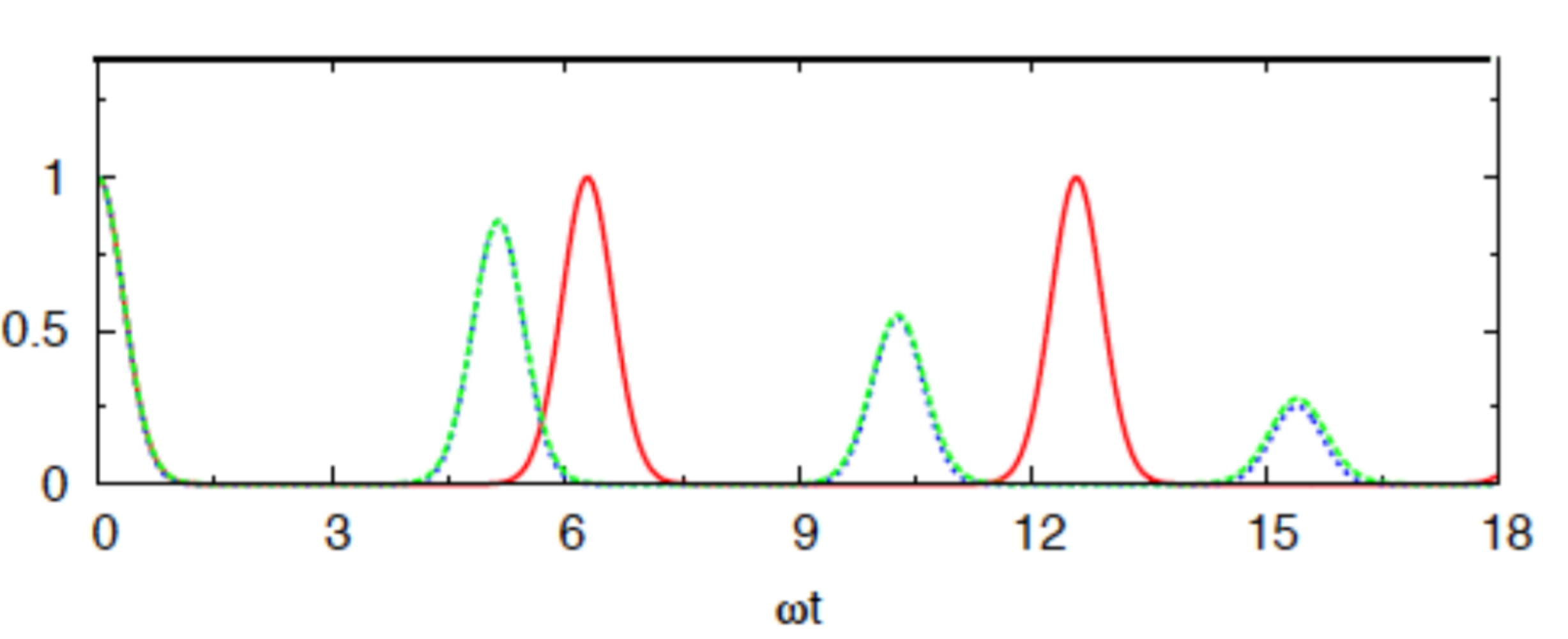}
\caption{ The magnitude of quantum coherence, 
$\vert \langle \uparrow \vert \rho_s(t) \vert \downarrow \rangle \vert$ for a central spin coupled to 200
bath spins with same frequency, $\omega_k = \omega$.  The coupling coefficients $g_k$ are sampled from a
uniform distribution.  The green, red, and blue-dashed curves correspond to the exact, second-order and modified (see text) second-order expansion of $\Gamma(t)$.}
\label{fig:purecoh}
\end{figure}

Secondly, when $\lambda_k \gg 1 $ then $\Omega_k$ is significantly shifted from 
the bare frequency $\omega_k$.  The density of states of the bath will be dramatically re-organized with respect to 
$\Omega_k$.
Any methods expanding around 
$\omega_k$ will be difficult to provide accruate results when system-bath coupling is strong and/or when 
$\omega_k$ is small.
A striking example would be a single-frequency bath, in which all modes possess the identical energy scale $\omega_k = \omega$ and coupled non-uniformly to the system.  For a bosonic bath as well as the second-order cumulant expansion for a spin bath, there is essentially no dephasing.  According to the second-order result in Eq.~(\ref{eq:ss_coherence}),
the coherence of the central spin is periodically recovered at time points $\omega t = n 2\pi$ with $n$ an integer.
However, for an exact treatment of the spin bath, non-trivial dephasing happens as long as not all coupling coefficients 
$g_k$ are identical.
Due to the system-bath interaction, the single-frequency distribution of $\omega$ can be broadened to a finite bandwidth corresponding to the 
dressed $\Omega_k$.  This analysis is confirmed in Fig.~\ref{fig:purecoh}, where the exact result (green) undergo
an irrversible decay but not for the second-order result (red).  In the figure, we also consider a modified 
second order cumulant result (blue dotted) which expands the decoherence function $\Gamma(t)$ with respect to $\Omega_k$ instead of $\omega_k$.  As shown, this modified expansion works extremely well.
In general, this dressing of $\Omega_k$ implies a faster dephasing rate is expected from a spin bath when compared
to a similar bosonic bath, i.e. a bath of harmonic oscillators sharing the same set of $\{\omega_k, g_k\}$.  

Finally, the temperature independence of $\Gamma(t)$ in Eq.~(\ref{eq:ss_coherence}) is 
another distinguishing property to set apart spin bath from what has been known for the bosonic bath models.  
This temperature-independent dephasing can already be inferred from the expression of the effective spectral density in Eq.~(\ref{eq:effspec}), and is further confirmed
to hold beyond the linear response regime in this pure dephasing model as the temperature factors is missing in the exact expression in Eq.~(\ref{eq:ss_coherence}).

To estimate the contributions of higher order cumulant corrections (HOCCs) to the central spin deocherence, 
we note the fourth order cumulants in the last line of Eq.~(\ref{eq:ss_coherence}) 
can become dominant under two conditions: (1) the perturbation parameters satisfy $\lambda_k > 1$ and/or 
(2) when $t > 1/(\lambda_k^2 \omega_k)$ such that the terms linearly proportional to $\omega_k t$ dominates the second order cumulant. The second condition implies a potential linear time $t$ divergence.  This instability is an artifact of cumulant expansion and can be removed by introducing higher order cumulants.
When the bath parameters $\{\omega_k, g_k\}$ are obtained by discretizing 
a continuous spectral density as discussed in Sec.~\ref{sec:2}, all $\lambda_k \sim 1/N_B$ can be made arbitrarily small when sufficiently large number of spin modes are used. Hence, the artificial divergence due to unstable part of the fourth-order cumulants can often be suppressed within the typical time domain for simulating condensed-phase dynamics.  However, even if just a few modes, satisfying $\lambda_k \geq 1$, 
could potentially contribute immensely to the overall HOCCs because of the logarithmic form for each spin's
contribution to the exact expression for $\Gamma(t)$ in Eq.~(\ref{eq:ss_coherence}).

\begin{figure*}
\centering
\subfloat{%
  \includegraphics[width=0.45\columnwidth]{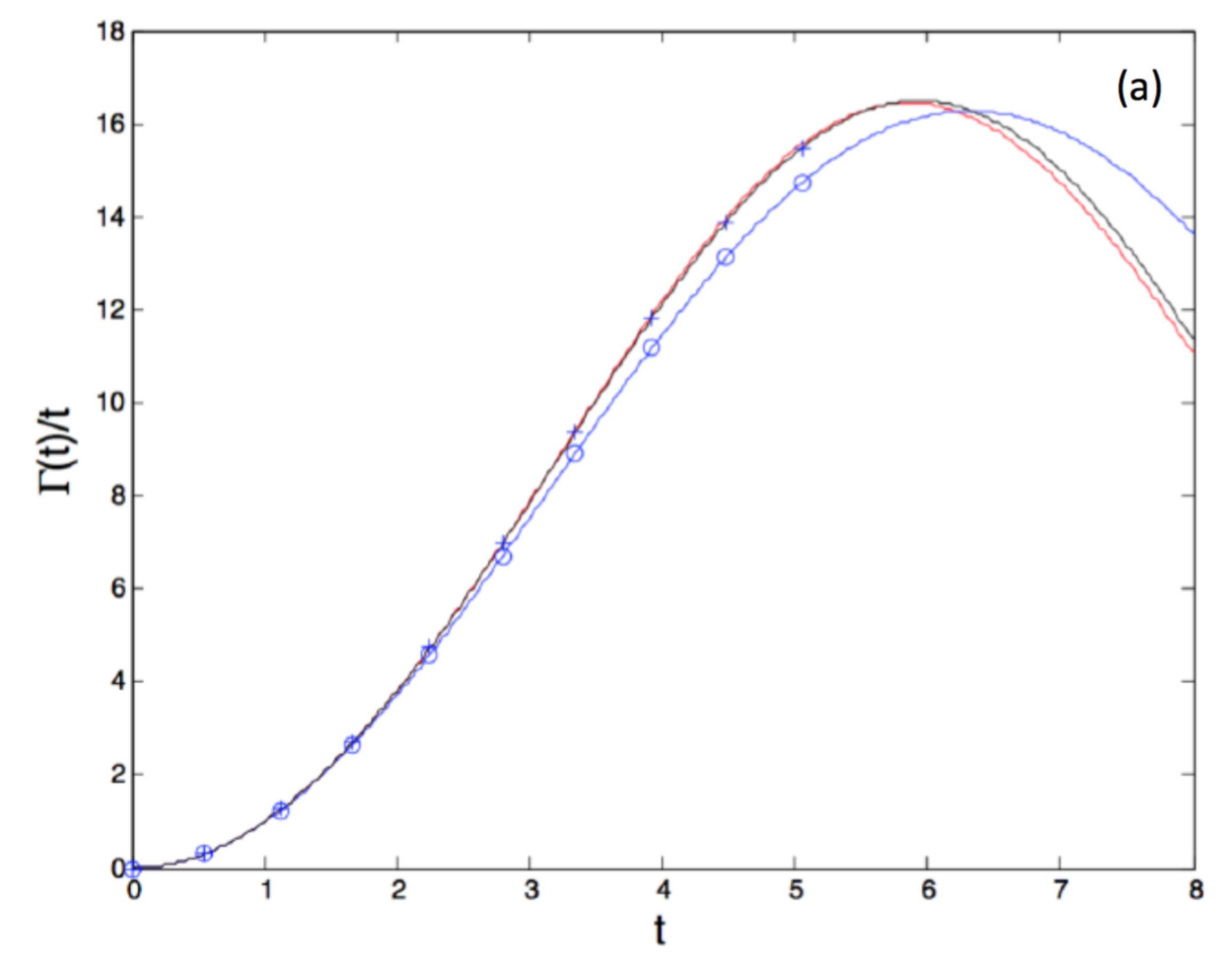}%
  \label{sfig:testa}
 }
\subfloat{%
  \includegraphics[width=0.45\columnwidth]{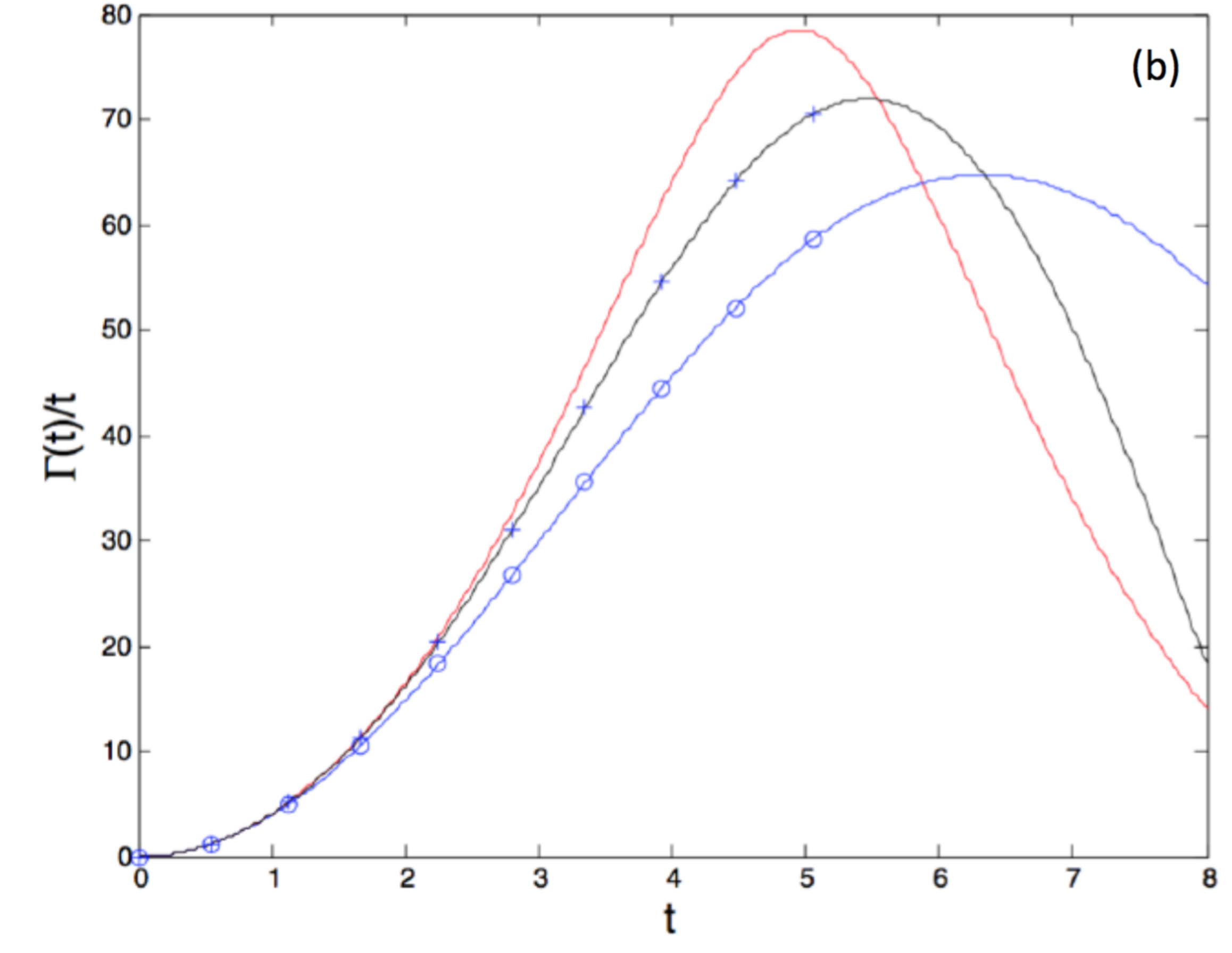}%
  \label{sfig:testb}
}
\caption{The decoherence rate $\Gamma(t)/t$ as function of time.
The red, black and blue curves are the exact rate, the fourth-order and
and the second-order rate according to Eq.~(\ref{eq:ss_coherence}).
The open circles on the curves are generated numerically from the gHEOM method.
The bath is composed of 50 spins with parameters $\omega_k$ and $g_k$ sampled
uniformly from the following ranges.
(Panel a): $\omega_k \in [0.4,0.6]$ and $g_k \in [0.08, 0.12]$.  
(Panel b): $\omega_k \in [0.4,0.6]$ and $g_k \in [0.18, 0.22]$.}
\label{fig:purerate}
\end{figure*}

As mentioned in the case of physical spin models\cite{Prokofiev:420342}, 
there is no reason that $\lambda_k$ should be related to the number of bath spins.  
Hence, the effects of HOCCs can become noticeable if not all $\lambda_k$ are sufficiently small within
the simulation time window to suppress the divergence associated with the unstable part of higher order cumulants. 
In Fig.~\ref{fig:purerate}, we look at the dephasing rate, $\Gamma(t)/t$ for two different cases to analyze HOCCs. The initial condition of the central spin is taken to be the pure state $\ket{\psi}=\frac{1}{\sqrt{2}}\left( \ket{\uparrow} + \ket{\downarrow} \right)$.  This calculation also serves as a benchmark to validate that 
the gHEOM correctly resolves the second and fourth
cumulant contributions when compared to the exact expansions in Eq.~(\ref{eq:ss_coherence}). 
The parameters $\epsilon=2$ and $\Delta=0$ are used for the system Hamiltonian.
We assign random samples of $\{\omega_k, g_k\}$ from two uniform 
distributions centered on $\omega_{o}$ and $g_o$, respectively, with details given in the figure caption.
In panel (a), the fourth order results can sufficiently reproduce the exact rate and it starts to deviate from the linear response rate around $t \sim 3$.  
In panel (b),  by further increasing the average coupling $g_0$, even the fourth order corrections starts to fall short of reproducing the exact rate around $t \sim 3$.  
For both cases considered in Fig.~\ref{fig:purerate},  $\lambda_k < 1$ hold for all bath modes, which ensure all
perturbative expansion parameters $\lambda_k$ are well-behaved in the short-time limit.
It is clear that the HOCCs becomes important to simulate the system relaxation when the spin bath 
parameters do not satisfy the scaling relation $\lambda_k \sim 1/N_B$.

\subsection{Anharmonic Condensed-Phase evironment}\label{sec:anharmonic-result}
We consider the same model as in Sec.~\ref{sec:dephasing} but with $\Delta \neq 0$ in Eq.~(\ref{eq:sysbath}).
The central spin could suffer relaxation due to interaction with the bath in this case.
The parameters,$\{\omega_k, g_k\}$, are assigned by discretizing an Ohmic spectral density.
This finite-size restriction is a necessity to observe any deviations from linear response results as explained
in the introduction.  Furthermore, the discretization allows us to numerically
compute the multi-time correlation functions, needed for the gHEOM calculations, 
by summing over contributions from each bath spins.

One can estimate the leading order corrections beyond the linear response approximation 
by a perturbative expansion with respect to $\lambda_k$ as 
done in the previous section. 
We first analyze the population dynamics in the Markovian limit.  
We adapt the NIBA (Non-Interacting Blip Approximation) equation to the spin bath model\cite{Segal:2014ws} with a
symmetric system Hamiltonian, $H_s = \frac{\Delta}{2}\sigma^x_0$. We find
\begin{eqnarray}\label{eq:niba}
\frac{ d}{dt} <\sigma^z_0(t)> = -\Delta^2 \int^t_0 ds e^{-Q_R(t-s)} \cos\left(Q_I(t-s)\right) <\sigma^z_0(s)>,
\end{eqnarray}
where the $Q_R$ and $Q_I$ functions read
\begin{eqnarray}\label{eq:qfuns}
Q_R(t) & \approx & \sum_k \left\{\lambda_k (1-\cos\left(\omega_k t\right)) + \frac{\lambda^2_k}{2}  
\left[\sin(\omega_k t)\omega_k t +
 \sin^2(\omega_k t) \sech^2\left(\frac{\beta\omega_k}{2}\right)  \right]
\right\}, \nonumber \\
Q_I(t) & \approx & \sum_k \left\{ \left[ \lambda_k \sin(\omega_k t) + 
\frac{\lambda_k^2}{2} \left[ \sin(2\omega_k t)-\cos(\omega_k t)\omega_k t \right] \right] 
\tanh\left(\frac{\beta\omega_k}{2}\right)  \right\},
\end{eqnarray}
with $\lambda_k \equiv (4g_k/\omega_k)^2$.
The equation (\ref{eq:niba}) is a second-order expansion (with respect to the off-diagonal element, $\Delta$) of 
the memory kernel. 
The functions, $Q_R(t)$ and $Q_I(t)$, in Eq.~(\ref{eq:qfuns}) are further expanded up to $\lambda_k^2$.
If one only retains the first term, proportional to $\lambda_k$,
then $Q_R(t)$ and $Q_I(t)$ reduce to the standard NIBA expressions for an effective bosonic bath
with a temperature dependent spectral density, Eq.~(\ref{eq:effspec}).

To finish the Markovian approximation, we replace $<\sigma^z_0(s)>$ with $<\sigma^z_0(t)>$ on the RHS
of Eq.~(\ref{eq:niba}), extend the integration limit to infinity in both directions, and perform a short-time expansions of $Q_R$ and $Q_I$ to keep terms
up to $\sim t^2$.  One then integrates out the memory kernel in Eq.~(\ref{eq:niba}) and obtains a simple rate equation with 
the Fermi golden rule rate given by
\begin{eqnarray}\label{eq:niba-rate}
k & = & \Delta^2 \sqrt{\frac{\pi}{a_1}}\left(1+\xi\right)^{-1/2}\exp\left(-\frac{b_1^2}{4a_1}\frac{(1+\zeta)^2}{1+\xi}\right), \nonumber \\
& \approx & \Delta^2 \sqrt{\frac{\pi}{a_1}}\exp\left(-\frac{b_1^2}{4a_1}\right)\exp\left(-\frac{1}{2}\xi\right), \nonumber \\
& = &  k_{\text{lin}}\exp\left(-\frac{1}{2}\xi\right)
\end{eqnarray}
where $\xi=a_2/a_1$, $\zeta=b_2/b_1$, $a_1 =\sum_k \lambda_k\omega_k^2/2$, $a_2 = \sum_k \lambda_k^2 \tanh(\beta\omega_k/2)\omega_k^2/2$, 
$b_1= \sum_k \lambda_k \omega_k \tanh(\beta\omega_k/2)$ and $b_2 = \sum_k \lambda_k^2 \omega_k \tanh(\beta\omega_k/2)/2$. As $\lambda_k \rightarrow 0$, the rate approaches to the linear response / effective bosonic bath results: $k \rightarrow k_{\text{lin}}$.  
The expression on the second and third line
constitutes a good approximation when both $\xi$ and $\zeta$ are small.  
In particular, the last exponential factor isolates the leading-order correction to the rate constant,
$\eta = \exp\left(-\frac{1}{2}\xi\right)$.
Whenever the scaling $\lambda_k \sim 1/N_B$ is imposed, $\xi$ will scale as $1/N_B$ too and $\eta$ will 
be exponenially suppressed.  Furthermore, we note that the leading-order correction reduces the relaxation rate at low temperature and gradually converge to the linear response result $k_{\text{lin}}$ with increasing temperature due to the factor $\tanh(\beta\omega_k/2)$ contained in $a_2$ variable.

\begin{figure*}
\centering
\includegraphics[width=0.9\linewidth]{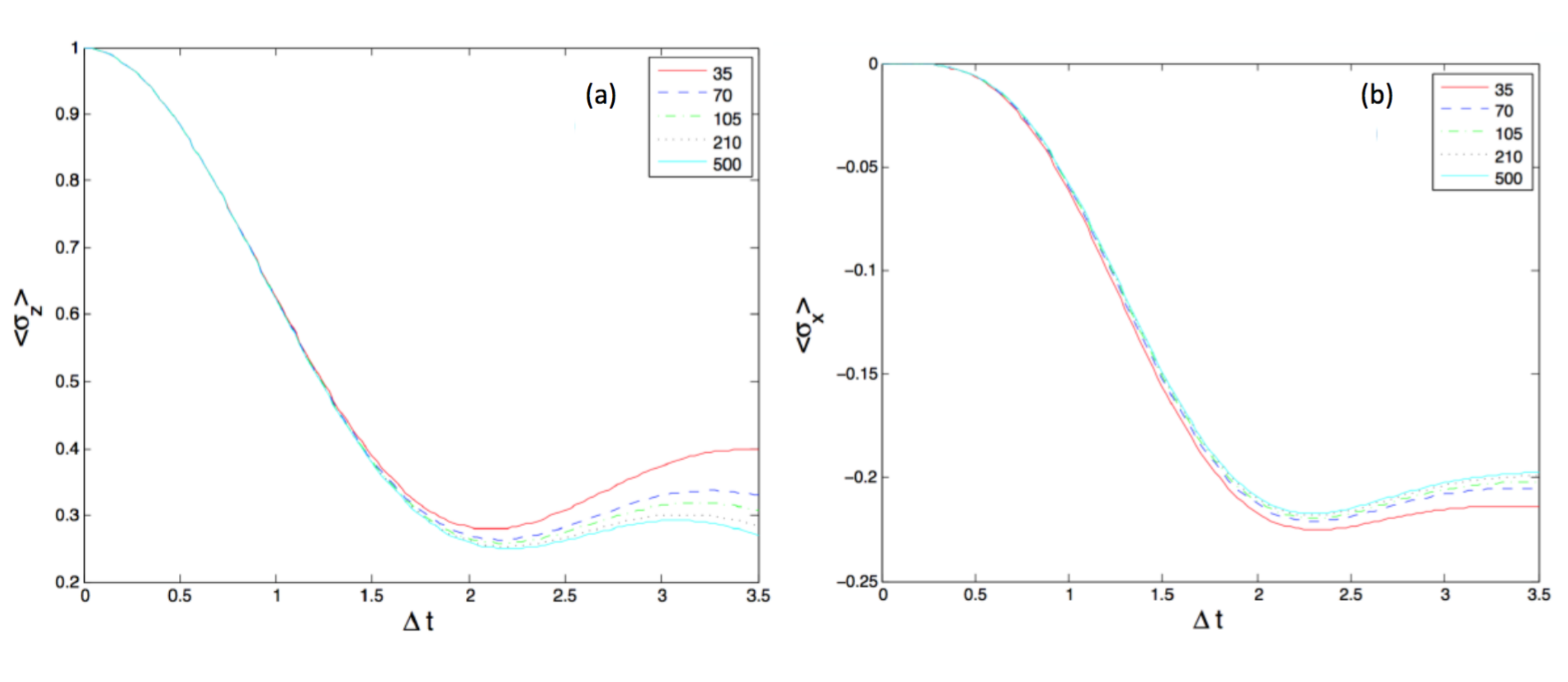}
\caption{Convergence study of the population and coherence dynamics with respect to the number of spins 
discretized from an Ohmic spectral density. 
The population (a) and coherence (b) dynamics are obtained with the fourth-order gHEOM.
For the 500-spin case, the fourth-order results essentially converge to the second-order results for both
population and coherence dynamics. 
}
\label{fig:num-ohmic}
\end{figure*}

Next, we investigate numerically the convergence of a spin bath to the linear response results within the gHEOM calculations.  
We use the following parameters
$\Delta=1$ and $\epsilon=0$ for the system Hamiltonian and 
the spectral density parameters: $\omega_c = \Delta$, $\beta \Delta = 2$,  
$\alpha = 2.3$.   For Ohmic spectral density, coupling coefficients satisfy $g_k^2 \propto \omega_k$ and
the perturbative parameter $\lambda_k \propto 1/\omega_k$.  Therefore,
the low frequency bath modes are more severely influenced by the
system-bath interactions with $\Omega_k=\omega_k\sqrt{1+\lambda_k}$ shifted further from $\omega_k$.
To investigate deviations from the linear response results, 
the highest frequency we consider is $\omega_{\text{max}} = 2 \omega_c$ in the discretization.
The convergence of dynamics is demonstrated in Figs.~\ref{fig:num-ohmic}a and \ref{fig:num-ohmic}b, the population
and the coherences of the central spin are plotted up to $\Delta t =  3.5$, respectively, for
a total number of $35$, $70$, $105$, $210$ and $500$ spins including up to the fourth-order cumulant expansions.  
As the number of bath spins increase, the results converge smoothly to those of the condensed phase environment in
the thermodynamic limit.  At $N_B=500$, the fourth order results already converge with the second order results.

\begin{figure}
\centering{
\label{fig:temp-ohmic}
\includegraphics[width=\linewidth]{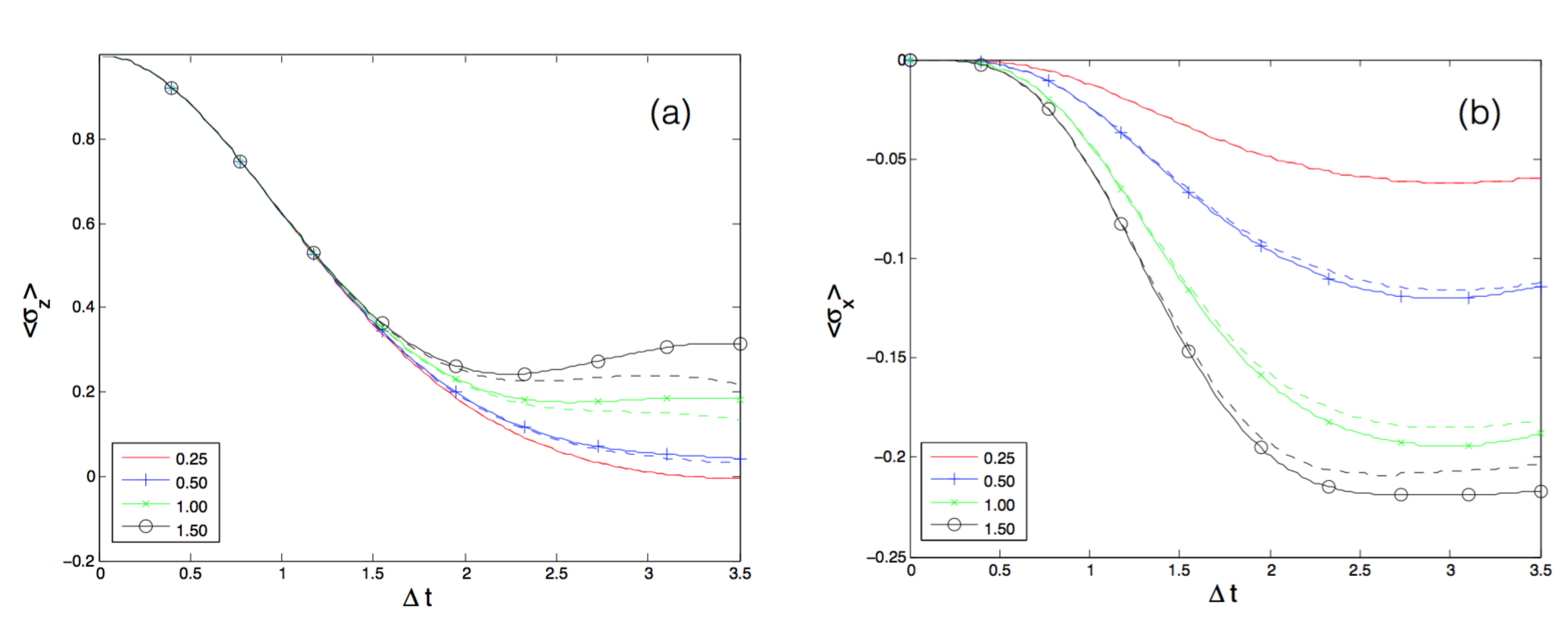} 
}
\caption{Temperature dependence of the population (a) and coherence (b) dynamics.
In both panels, four temperature cases $\beta\Delta = 0.25$ (red), $\beta\Delta = 0.5$ (blue,$+$), $\beta\Delta = 1.0$ (green, $\times$) and $\beta\Delta = 1.5$ (black, $o$) are considered.
The solid and dashed curves correspond to the second-order and fourth-order gHEOM calculations, respectively.}
\end{figure}

Next, we investigate the temperature dependence of HOCCs. 
The same set of Hamiltonain and bath's spectral density parameters as above is 
used except the temperature will be varied for analyses.
We use a set of $N_B=35$ spins for illustrations.  
In many earlier studies, distinctive properties of a spin bath (in comparison to a bosonic counterpart)
are found to be temperature-related and are attributed to the temperature-dependent
spectral density, Eq.~(\ref{eq:effspec}).  
However, restricting the discussions to the effective spectral density imply the comparisons focused on
the linear response limit.  
We would like to further analyze the contributions of HOCCs to these temperature-dependent effects.
In Fig.~\ref{fig:temp-ohmic}, we compare the second-order (dashed curves) and fourth-order (solid curves) results to inspect the contribution of the HOCCs.
In short, the HOCCs become more pronounced when the temperature is lowered and the divergence between second-order and four-order results increase.  The numerical results is also consistent with the earlier 
conclusion drawn from the rate expression, Eq.~(\ref{eq:niba-rate}).
Similar observations\cite{makri99,Wang:2012kk} have been reported in the literature where it was found that more number of discretized bath spins are needed to reach the linear response limit at low temperatures. Despite the simple argument that the linear response of a spin and a harmonic oscillator converge in
the zero temperature limit, the two bath models actually do not converge except in the cases of large bath size. This is because the higher order response functions for spins become more prominent in the low temperature regime. The primary reason is due to the way temperature enters the correlation functions as $\tanh(\beta\omega_k/2)$.

\subsection{Ising Spin Bath}\label{sec:res-phys}
Finally, we consider another system-bath interaction, $\hat H_{\text{int}} = \hat \sigma_0^z \sum_k g_k 
\hat \sigma_k^z$.  The Ising spin-spin interaction prohibits bath spins to be flipped but still entangles system 
and bath. While this spin bath model is appropriate in certain quantum computing contexts\cite{Krovi:2007iu,torron_kosloff_njp16}, it does not relate
to a condensed phase environment. Hence, we will follow the physical spin bath approach to sample $\omega_k$ and
$g_k$ from uniform distributions.

We first consider a pure dephasing case with $\Delta=0$.  The coherence
of the central spin can be cast in the general form of Eq.~(\ref{eq:off-diagonal}) but with a different
decoherence function,
\begin{eqnarray}\label{eq:off-diagonal2}
\Gamma(t) & = & \sum_k \ln\left(\cos(2g_k t) - i \gamma_k \sin(2g_k t) \right),
\end{eqnarray}
where $\gamma_k = \tanh(\beta\omega_k/2)$.  Unlike the previouse pure-dephasing model in Sec.~\ref{sec:dephasing},
$\Gamma(t)$ acquires a temperature dependence through $\gamma_k$. 
Since $\hat H_B$ and $\hat H_{\text{int}}$ commutes, the
bath Hamiltonian can be removed from the dynamical equation in a rotated frame and no dressed bath frequencies
$\Omega_k$ appear as in Sec.~\ref{sec:dephasing}.  For the Ising spin bath, the bath frequencies $\omega_k$ enter
the decoherence function through $\gamma_k$, reflecting the
thermal equilibrium initial condition $\rho^{eq}_B$.
	
According to Eq.~(\ref{eq:off-diagonal2}), $\gamma_k$ modulate the magnitude of $\Gamma(t)$ and 
the Ising spin bath becomes less efficient at 
interacting with the system in the high temperature limit and / or the low-frequency limit when $\gamma_k \ll 1$.
The system-bath coupling coefficients $g_k$ determine the oscillatory behaviors of $\Gamma(t)$ 
in time domain. When $g_k$ are narrowly distributed around a mean value $g_o$, one expects a partially periodic recurrence of $\Gamma(t)$.  

\begin{figure}
\centering
\includegraphics[width=\linewidth]{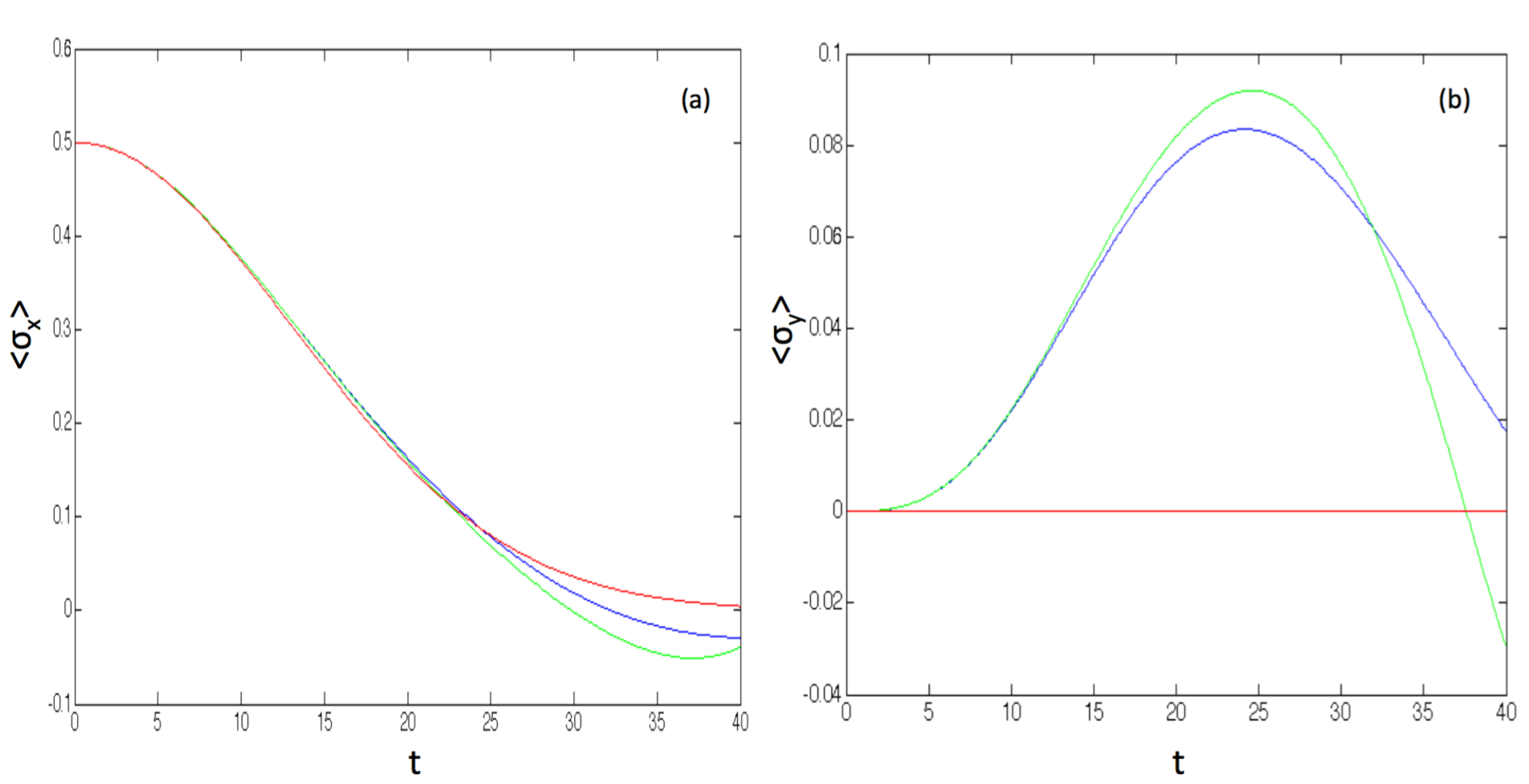}
\caption{The real (a) and imaginary (b) part of coherence, $\langle \uparrow \vert \rho_s(t) \vert \downarrow \rangle$, for a central
spin coupled to 50 bath spins.  The system Hamiltonian is absent, i.e. $\epsilon =0$ and $\Delta=0$. The
bath parameters are randomly drawn from the range: $\omega_k \in [14,15]$ and $g_k \in [0.006,0.0018]$ and
$\beta=0.2$.}
\label{fig:norotation}
\end{figure}

\begin{figure}
\centering
\includegraphics[width=\linewidth]{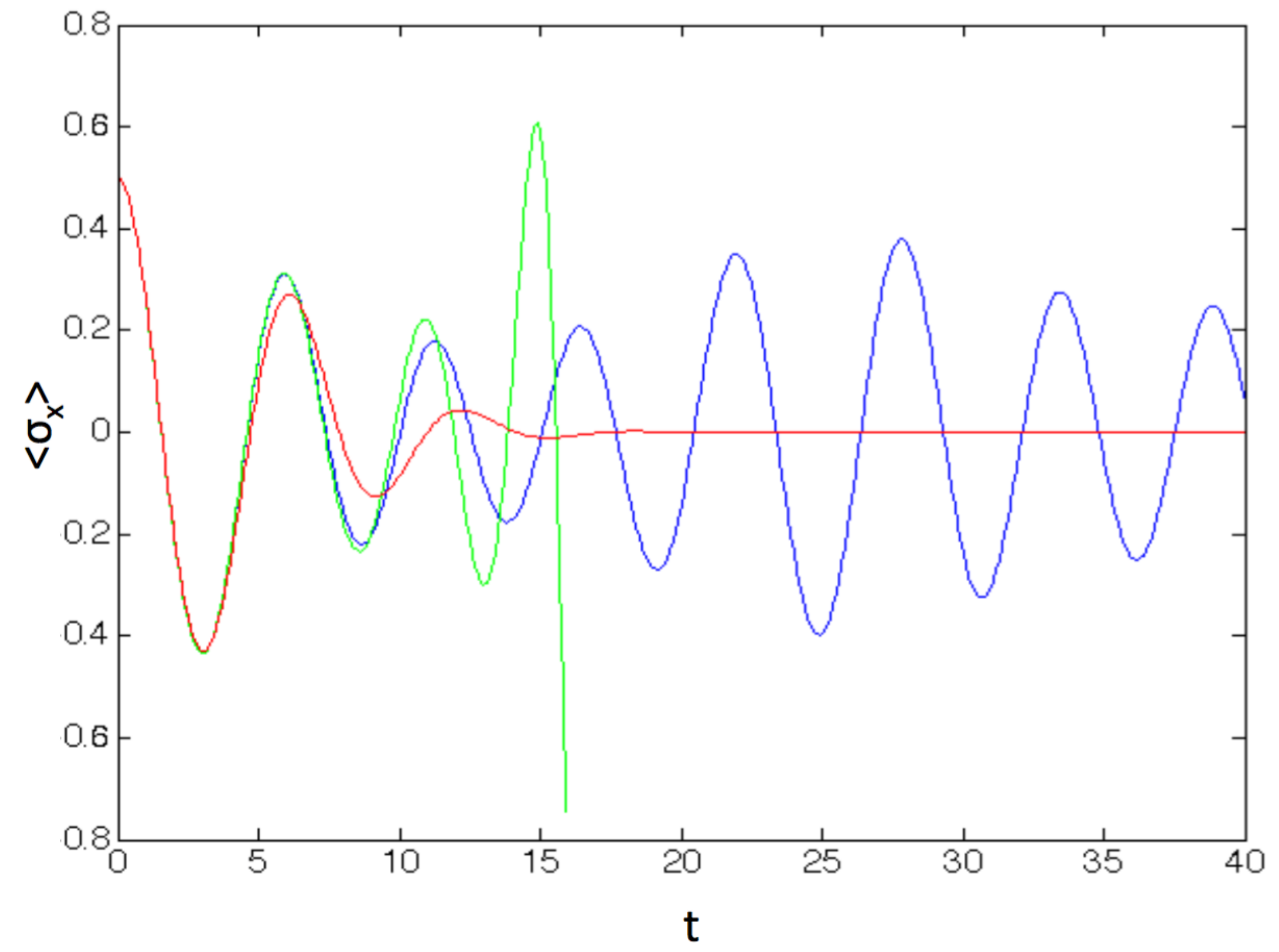}
\caption{The real part of coherence, $\langle \uparrow \vert \rho_s(t) \vert \downarrow \rangle$, for a central
spin coupled to 30 bath spins.  The system Hamiltonian parameters are $\epsilon =1$ and $\Delta=0$. The
bath parameters are randomly drawn from the range: $\omega_k \in [7.8,8.1]$ and $g_k \in [0.005,0.007]$ and
$\beta=0.5$.}
\label{fig:nodecay}
\end{figure}

Now we investigate effects of HOCCs. Two points make the Ising spin model an interesting case to analyze.  
First, an expansion of Eq.~(\ref{eq:off-diagonal2}) with respect to $g_k$ reveals that the even cumulants do not contribute to the imaginary component of $\Gamma(t)$, which drives a rotation of the central spin on the Bloch sphere. A second order cumulant expansion will completely miss this rotation.
This point is illustrated in Fig.~\ref{fig:norotation} where the central spin's initial condition is taken to be
$\ket{\psi_s(0)} = (\ket{\uparrow}+\ket{\downarrow})/\sqrt{2}$.
The second-order result (red curve) reproduces the real part of the coherence, $\langle\uparrow \vert \rho_{s}(t) \vert \downarrow \rangle$, but completely misses the growth of the imaginary component.  
Once we incorporate the third and fourth cumulants in the calculation, the result (green curve) converges better to the exact one in Fig.~\ref{fig:norotation}b.
Secondly, take $B=\sum_{k} g_k \hat\sigma_k^z$ as the bath part of $\hat{H}_{\text{int}}$ and it is easy to verify 
the multi-time correlation functions, such as 
$\text{Tr}\left\{\rho^{eq}_B B(t_1)B(t_2)\dots B(t_n)\right\}$, are time invariant (i.e. the memory kernel of the bath do not decay in time).  
The highly non-Markovian nature of the Ising spin bath can give rise to non-trivial steady states in the long time limit.  In Fig.~\ref{fig:nodecay}, we consider another case and find the second-order result suggests a fully
decayed coherence while the exact result shows a persistent oscillation.  Although the higher-order result (green curve) can better capture the on-going oscillating behavior in the transient regime, the artificial divergence of cumulant expansion (explained in Sec.~\ref{sec:dephasing}) suggests more cumulant terms should be included within the simulated time domain.

\begin{figure}
\centering
\includegraphics[width=\linewidth]{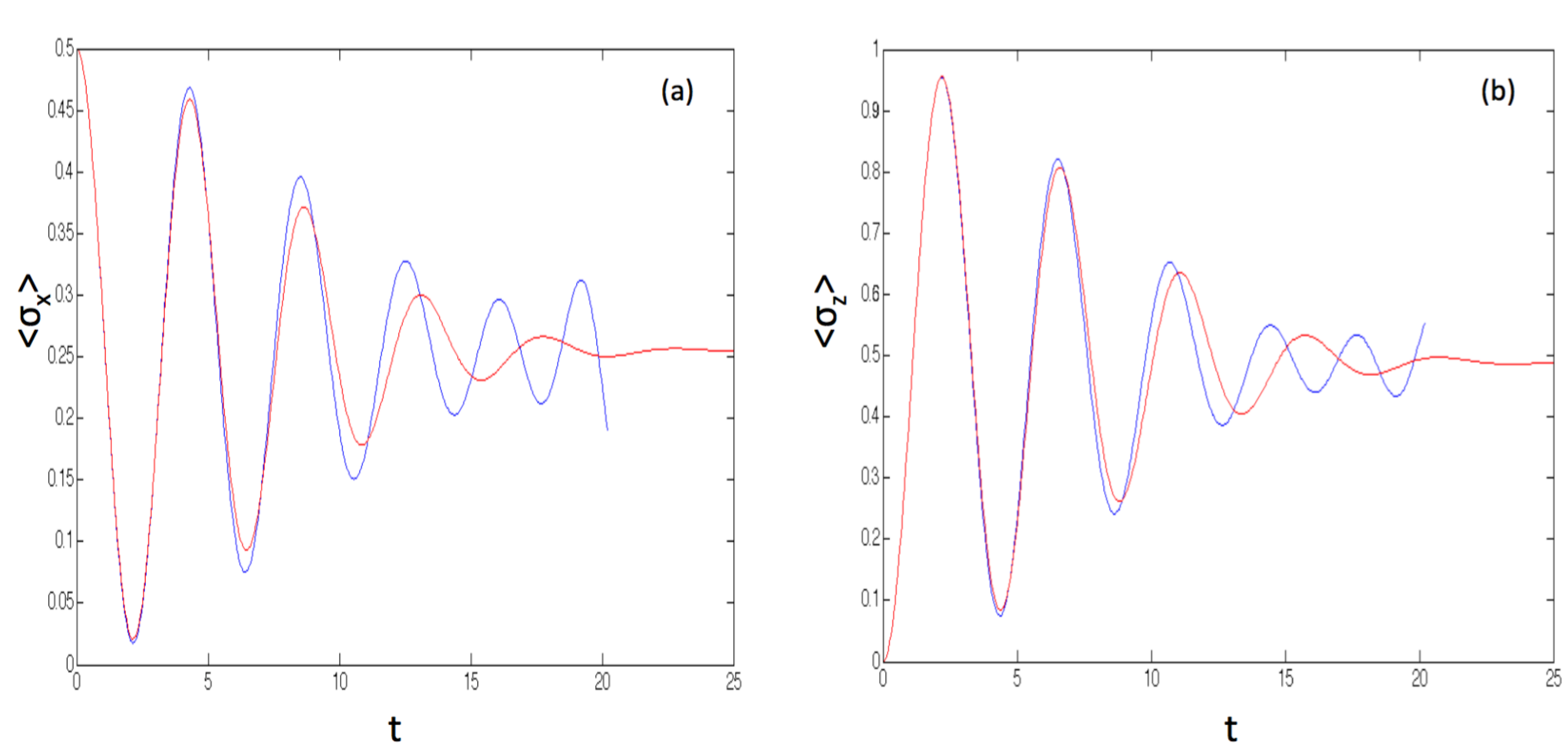}
\caption{The coherence (a) and population (b) dynamics for a central
spin coupled to 45 bath spins.  The system Hamiltonian parameters are $\epsilon =1$ and $\Delta=1$. The
bath parameters are randomly drawn from the range: $\omega_k \in [6.8,7.2]$ and $g_k \in [0.04,0.06]$ and
$\beta=0.5$.}
\label{fig:last}
\end{figure}

Going beyond the pure dephasing case, we restore the off-diagonal coupling $\Delta$.  We examine both
the coherence and population dynamics in Fig.~\ref{fig:last}.  Similar to pure dephasing case, one expects a slow decoherence when $g_k$ are narrowly distributed.  For both coherence and population dynamics, we identify the clear insufficiency of second-order results and higher-order cumulants again helps to restore the coherence and the beatings of the population dynamics in the transient regime.

\section{Conclusions}\label{sec:Con}

In summary, we use our recently proposed gHEOM to investigate the effects of HOCCs on 
the quantum dissipations induced by finite-size spin bath models.  
The gHEOM can systematically incorporate
the higher order cumulants of the bath's influence functional into
calculations. The controlled access to non-Gaussian effects of the bath   
allows us to assess the sufficiency of a linear response approximation.
Besides the spin baths, the methodology can be similarly applied towards other types of 
anharmonic environments.  However, due to the prohibitive
numerical resources required to accurately characterize the high-dimensional
multi-time correlations functions, the best usage of this method
is to combine it with transfer tensor method (TTM) proposed by one of us. 
One can use the gHEOM to quantitatively capture 
the exact short-time dissipative dynamics embedded in a non-Gaussian bath.  
These short-time results are then fed to the TTM method to reproduce the correct memory kernel of the 
environment and allow an efficient and stable long-time simulation of dissipative dynamics.

Through the analyses done in Sec.~\ref{sec:anharmonic-result}, we find the linear response approximation provides
a highly efficient and accurate result for a finite spin bath over a wide range of parameters.  
This is mainly because the next leading order correction scale as $1/N_B$ in the cumulant expansion.
We present one ``extreme" result for a relatively slow Ohmic bath
in order to observe appreciable corrections coming from the higher order cumulant terms in the short time limit.  Even in this case, the higher order effects still vanish when $N_B = 500$. Although, the low-temperature condition should exacerbate the discrepancy between exact and linear-response results, we find the actual effects rather minimal
in the short-time limit. Considering the significant numerical costs to access higher order cumulants, it 
certainly make linear response approximation a highly appealing option in dealing with a spin-based condensed
phase environment. In App.~\ref{app:linear-resp}, we further investigate the differences between a spin and bosonic
bath in the linear response limit.  We confirm the lack of appreciable temperature dependence on the dissipative dynamics and the emergence of negative differential thermal conductance are two robust physical signatures to distinguish a spin bath from a corresponding bosonic one as explained in the appendix.

It is much simpler to devise numerical examples in which the higher order cumulants
play critical roles in a physical spin bath model. The most critical factor is the probability
distributions for $\{\omega_k, g_k\}$.  Narrow distributions will make the spin bath model more difficult for 
linear-response approximations, and it is likely to have such narrow distributions in real spin-based environments.
In such cases, linear-response results could deviate extremely from the exact results such as shown in Fig.~\ref{fig:purecoh} and Fig.~\ref{fig:nodecay}.
Secondly, for highly symmetric spin-spin interaction such as the Ising Hamiltonian considered in 
Sec.~\ref{sec:res-phys}, the second order cumulants fail to generate a rotation of the spin state, which could only
be accounted by the odd-order cumulants.  Finally, the physical spin bath could be difficult to handle due to the possibility of extreme non-Markovianity.  In the Ising Hamiltonian example, we see the extreme case of having 
all bath's multi-time correlation functions to be time-invariant and we need to expand deep down the hierarchy to obtain converged results. The highly non-Markovian nature of spin bath is not a rare exception.  In addition to Ising Hamiltonian, the flip-flop and Heisenberg Hamiltonian in combination with narrow distribution of $\{\omega_k, g_k \}$ can also result in highly symmetric systems (central spin plus the bath) 
with a rich set of non-Markovian and persistent dynamics.
With the gHEOM method, we can systematically incorporate higher order cumulants to improve the simulation results for physical spin bath models.

\begin{acknowledgements}
C.H. acknolwedges support from the SUTD-MIT program.  J.C. is supported by NSF (grant no. CHE-1112825) and SMART.
\end{acknowledgements}

\appendix

\section{Stochastic mean field and multi-time correlation functions}\label{app:stoch}
From Eq.~(\ref{eq:bfield}), it is clear that equation of motions for $\mathcal{B}(t)$ can be obtained from
the stochastic dynamical equations for the bath density matrix in Eq.~(\ref{eq:ito2}).
More specifically, the time evolution of the stochastic field $\mathcal{B} = \sum_k g_k (\langle b^{\dag}_{k} \rangle + \langle b_{k} \rangle)$ 
is jointly determined by 
\begin{eqnarray}
\label{eq:boson_mode1}
d\langle b^{\dag}_{k} \rangle & = & 
i\omega_{k} \langle b^{\dag}_{k} \rangle dt
+ \frac{1}{\sqrt 2} g_{k} dW^* \mathcal{G}^{+-}_k  
+ \frac{1}{\sqrt 2} g_{k} dV^* \mathcal{G}^{-+}_k,\\
d \langle b_{k} \rangle & = & -i \omega_{k} \langle b_{k} \rangle dt
+\frac{1}{\sqrt 2} g_{k} dW^* \mathcal{G}^{-+}_k 
+ \frac{1}{\sqrt 2} g_{k} dV^* \mathcal{G}^{+-}_k. 
\label{eq:boson_mode2}
\end{eqnarray}
The expectation values in Eqs.~(\ref{eq:boson_mode1})-(\ref{eq:boson_mode2}) are taken with respect to the stochastically evolved 
$\tilde\rho_k(t)$.  The generalized cumulants above are defined as
\begin{eqnarray}\label{eq:2ndcum}
\mathcal{G}^{\alpha_1 \alpha_2}_k = \langle b^{\alpha_1}_k b^{\alpha_2}_k \rangle - \langle b^{\alpha_1}_k \rangle  \langle b^{\alpha_2}_k \rangle.
\end{eqnarray}
In this case of bosonic bath models, it is straightforward to show that the time derivatives
of $\mathcal{G}^{\alpha_1 \alpha_2}_k$ vanish exactly.  Hence, the second order cumulants are determined by the thermal equilibrium conditions
of the initial states.  Immediately, one can identify the relevant quantity $\mathcal{G}^{+-}_k = n_{B}(\omega_k)$, 
the Bose-Einstein distribution for the thermal state of the bath.   
The time invariance of the second order cumulants make Eq.~(\ref{eq:boson_mode1})-(\ref{eq:boson_mode2}) amenable to deriving a closed form solution.
On the other hand, for non-Gaussian bath such as the spin bath,  
the second cumulants are not time invariant. One way to determine their 
time evolution is to work out their equations of motion by iteratively applying
Eq.~(\ref{eq:ito2}).  It is straightforward to show these equations couples different orders of generalized cumulants, 
\begin{eqnarray}\label{eq:gencum}
d \mathcal{G}^{\pmb \alpha}_k & = &
 i dt \vert \pmb{\alpha} \vert \omega_k \mathcal{G}^{\pmb \alpha}_{k}
+ \frac{1}{\sqrt{2}} dW^*_s \left(\mathcal{G}^{[\pmb{\alpha},+]}_{k}+\mathcal{G}^{[\pmb{\alpha},-]}_{k}\right) \nonumber \\
& & + \frac{1}{\sqrt{2}}dV^*_s \left(\mathcal{G}^{[+,\pmb{\alpha}]}_{k}+\mathcal{G}^{[-,\pmb{\alpha}]}_{k}\right),
\end{eqnarray}
where $\pmb \alpha = (\alpha_1, \alpha_2 \dots \alpha_n)$ specifies a sequence of raising and lowering spin operators that constitute this
particular $n$-th order cumulant and $\vert  \pmb\alpha \vert = \sum_i \alpha_i$ with $\alpha_i = \pm 1$ depending on whether it refers
to a raising (+) or lowering (-) operator, respectively.   
We use $[\pmb \alpha, \pm] \equiv (\alpha_1 , \dots \alpha_n, \pm)$ to denote an $n+1$-th cumulant obtained by appending
a spin operator to $\pmb \alpha$.  A similar definition is implied for $[\pm, \pmb \alpha]$.  More specifically, 
these cumulants are defined via an inductive relation that we explicitly demonstrate with an example to obtain a 
third-order cumulant starting from a second-order one given in Eq.~(\ref{eq:2ndcum}),
\begin{eqnarray}
\mathcal{G}^{[(\alpha_1,\alpha_2),\pm]}_k = \langle b^{\alpha_1} b^{\alpha_2} (b^\pm - \langle b^\pm \rangle) \rangle + 
\langle b^{\alpha_1}(b^\pm - \langle b^\pm \rangle) \rangle \langle b^{\alpha_2}\rangle +
\langle b^{\alpha_1} \rangle \langle b^{\alpha_2} (b^\pm - \langle b^\pm \rangle) \rangle.
\end{eqnarray} 
The key step in this inductive procedure is to insert an operator identity $b^\pm - \langle b^\pm \rangle$ at the end of each expectation bracket 
defining the $n$-th cumulant. If a term is composed of m expectation brackets, then this insertion should apply to one bracket at a time and generate
m terms for the $n+1$-th cumulant.
Similarly, we get $\mathcal{G}^{[\pm,\pmb \alpha]}$ by inserting the same operator identity to the beginning of each expectation bracket of 
$\mathcal{G}^{\pmb \alpha}_k$.

For the spin bath, these higher order cumulants persists up to all orders.
In any calculations, one should certainly truncate the cumulants at a specific $k$-th order by imposing the time invariance, 
$G^{\pmb \alpha}_k(t) = G^{\pmb \alpha}_k(0)$ and evaluate the lower order cumulants by recursively integrating
Eq.~(\ref{eq:gencum}).  Through this simple prescription, one derives Eq.~(\ref{eq:generalfield}).



\section{Physical signatures of the spin bath models in the thermodynamic limit}\label{app:linear-resp} 
In the main text, we focus predominantly on the higher order corrections
to the quantum dissipations induced by a spin bath.  
Now we turn attention to the linear response limit, and we 
look for physical signatures that can distinguish the spin and bosonic bath models . 
The condensed-phase spin bath (with practically infinite number of modes) 
can be rigorously mapped onto an effective bosonic bath with a 
temperature-dependent spectral density, Eq.~(\ref{eq:effspec}).  
To facilitate the calculations in this appendix, 
we should take the spin bath as an effective bosonic model and 
adopt the superohmic spectral density for convenience.
The results below compliments those of earlier studies\cite{Segal:2014ws,Wang:2012kk,Gelman:2004ji,Zhang:2015jl,Lu:2009hd} on the same subject. 

\subsection{Electronic Coherence of a Two-Level System}\label{app:pop}

We first compare the dynamics of a dimer coupled to (1) a bosonic bath and (2) a spin bath.
The spectral densities for the two models read
\begin{eqnarray}\label{eq:appspec}
J_b(\omega) & = & 2 K \frac{\omega^3}{\omega_{c}^2}e^{-\omega/\omega_{c}}, \nonumber \\
J_s(\omega) & = & 2 K \frac{\omega^3}{\omega_{c}^2}e^{-\omega/\omega_{c}} \tanh\left(\beta\omega/2\right),
\end{eqnarray}
where the subscript $b/s$ denotes the bosonic bath and the spin (effective bosonic) bath model, respectively.
In the study of FMO-like molecular systems, surprisingly long coherence times were discovered and successfully explained through an enhanced NIBA formalism\cite{weiss:book} which takes
into account the first order blip interactions.  By considering the same dimer system and a similar set of experimentally relevant parameters as in Ref.~\onlinecite{Pachon:2011ea}, 
we investigate how much the coherent population dynamics of an effective dimer will change when the environment is replace by a spin bath with an identical set of parameters as the original harmonics-oscillator based condensed phase. 
The enhanced NIBA formalism is also adopted here as it provides sufficiently accurate results over a long time span in the parameter regime considered here. 
The parameters are $K=0.16$, $\epsilon/\Delta=0.6$, $\omega_c /\Delta = 2.0$ and $\Delta \approx 106.2 \text{cm}^{-1}$.  

\begin{figure*}
\centering
\subfloat{%
  \includegraphics[width=0.45\columnwidth]{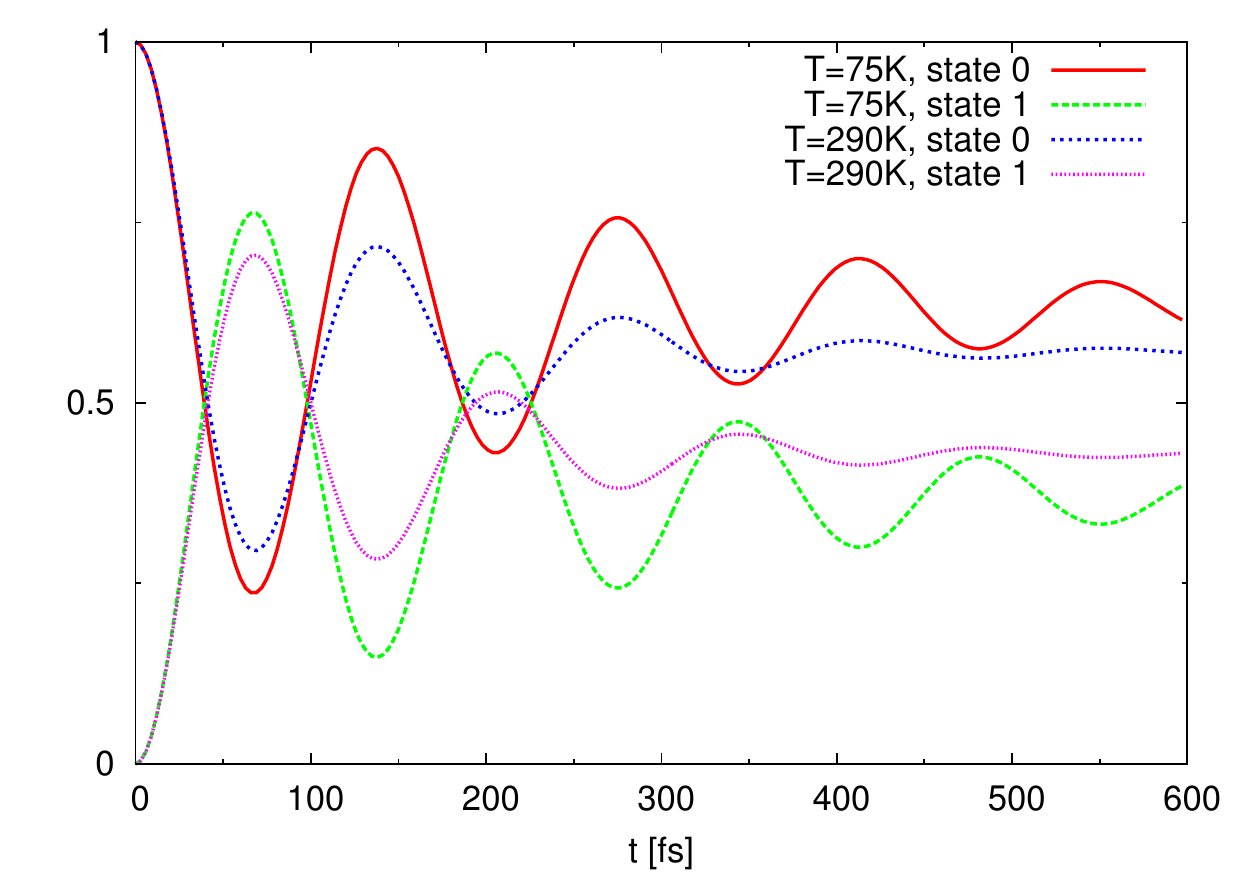}%
  \label{sfig:app-pop1}
}
\subfloat{%
  \includegraphics[width=0.45\columnwidth]{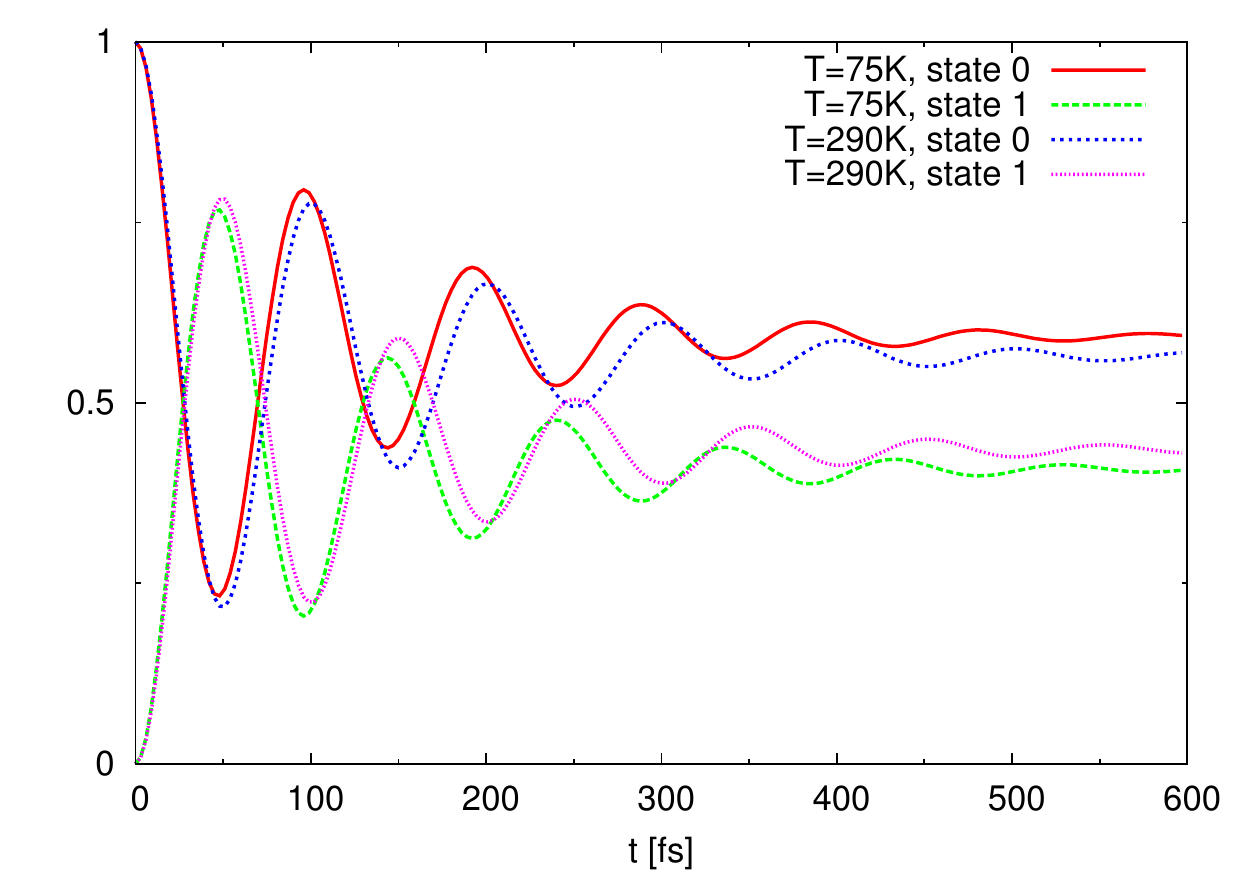}%
  \label{sfig:app-pop2}
}
\caption{Relaxation in the presence of a bosonic bath (left) and a spin bath (right) at high and low temperatures.
}
\label{fig:app-pop}
\end{figure*}

From Fig.~\ref{fig:app-pop}, the dimer's population dynamics in the presence of the bosonic (panel a)
and the spin (panel b) bath at two different temperatures: $T=290 K$ and $75 K$ are presented.
The significant temperature-dependent relaxation is observed in the standard bosonic bath; while
the almost temperature independent relaxation is found in the spin bath model.
These results confirm that the atypical temperature dependence of a spin bath induced relaxation can 
be quite pronounced and easily detected in the experimentally accessible regime.  


\subsection{Energy transport through a non-equilibrium junction}
We next explore additional features of a spin-based environment in the non-equilibrium situations.  
We study the energy transport across a molecular junction connected to two heat baths composed of non-interacting
spins. The extended double-bath model should read,
\begin{eqnarray}
\hat{H} = \hat{H}_s + \sum_{\mu=L,R} \hat{H}_{B,\mu} + \sum_{k,\mu=L,R} \hat{\sigma}^z_0 (g_{k\mu} \sigma^{\dag}_{k,\mu} + g_{k\mu} \sigma_{k,\mu}),
\end{eqnarray}
where $\hat{H}_s$ and $\hat{H}_{B,\mu}$ are the standard system and bath Hamiltonian with $\mu=L,R$ to denote the two bath at left or right end.   
Recent study\cite{Segal:2014ws} tried to model anharmonic junctions with the spin bath and  hinted several qualitative differences in the transport phenomena.  In this work, we adopt a non-equilibrium Polaron-transformed Redfield equation (NE-PTRE) in conjunction with the full counting statistics to compute the steady-state energy transfer through the junction.  
In Ref.\onlinecite{Wang:2015jz}, it was demonstrated that NE-PTRE can be reliably used to calculate energy currents (through a 
molecular junction coupled to bosonic baths) from the weak to the strong system-bath coupling regimes as the corresponding analytical expressions for the energy 
current reduced elegantly to either the standard Redfield (weak coupling) or NIBA (strong coupling) results in the appropriate limits.  
In this study we apply the NE-PTRE formalism to calculate and compare the energy current through the junction while contacted by (a) two spin baths and (b) two bosonic baths held at different temperatures. 
Similar to App.~\ref{app:pop}, the actual calculations below will treat the spin bath as an effective bosonic bath.  
The same superohmic spectral density, Eq.~(\ref{eq:appspec}), is adopted here.  
The parameters are $\Delta=1$, $\epsilon=10$, $\omega_c=10$ and $K=3.5$. 
The left and right bath share an identical set of parameters except the temperature, and the fast bath (or scaling) limit with $\omega_{c} \gg 1$ is imposed in both bath models.  
\begin{figure}
\begin{center}
\includegraphics[width=0.9\textwidth]{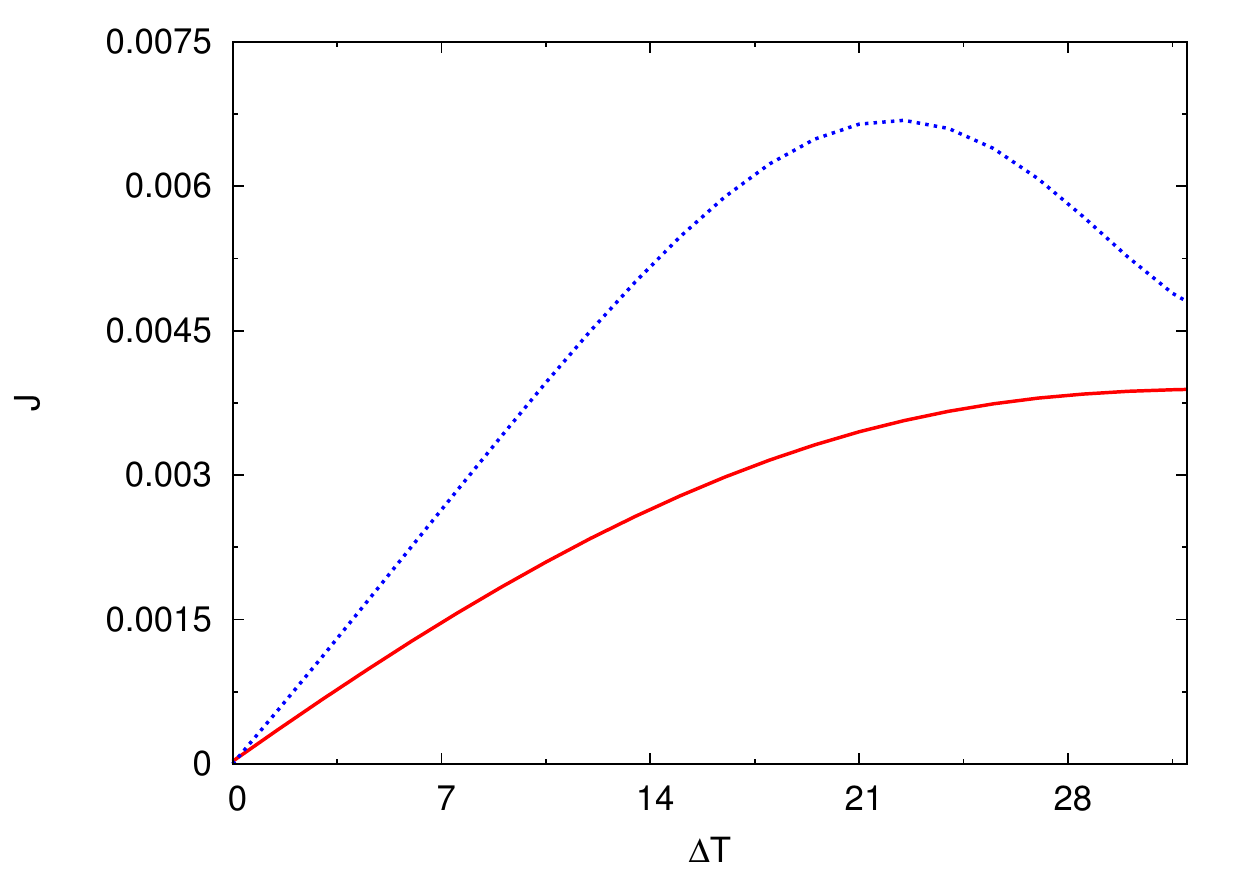}
\end{center}
\caption{The steady-state energy current through a two-level junction as a function of the temperature difference between the left and right baths. 
The energy current for the spin bath (blue,dotted) and bosonic bath (red,solid) in the strong coupling regimes, respectively.
The temperature increment $\Delta T =  T_L - T_R$ is applied symmetrically to both baths such that $T_L + T_R$ is fixed. 
}
\label{fig:ndtc}
\end{figure}

In Fig.~\ref{fig:ndtc}, the heat transfer exhibits a negative differential thermal conductance (NDTC) for spin baths but not for bosonic baths.
The temperature, $k_B T_\alpha$ are measured in terms of $\Delta$ with $k_B=1$ in the present case. 
We remark that the NDTC (for bosonic bath models) reported in some earlier studies are found to be an artifact of
the Marcus approximation (over a wide range of parameter regimes) as clarified by the NE-PTRE method reported in 
Ref.~\onlinecite{Wang:2015jz}.  In Fig.~\ref{fig:ndtc}, the NE-PTRE predicts the NDTC for the spin bath model but not for
the bosonic bath model.  In a separate work (not shown here), we also investigate a single-frequency spin bath and 
compute the energy current by using the NIBA method and find the same NDTC appearing in the strong coupling regime.  This helps to validate that the present result is not specific to a 
particular spectral density.
This qualitative difference between the bosonic and spin baths results make the NDTC another strong physical signature to distinguish the two bath models.


\begin{thebibliography}{44}
\expandafter\ifx\csname natexlab\endcsname\relax\def\natexlab#1{#1}\fi
\expandafter\ifx\csname bibnamefont\endcsname\relax
  \def\bibnamefont#1{#1}\fi
\expandafter\ifx\csname bibfnamefont\endcsname\relax
  \def\bibfnamefont#1{#1}\fi
\expandafter\ifx\csname citenamefont\endcsname\relax
  \def\citenamefont#1{#1}\fi
\expandafter\ifx\csname url\endcsname\relax
  \def\url#1{\texttt{#1}}\fi
\expandafter\ifx\csname urlprefix\endcsname\relax\def\urlprefix{URL }\fi
\providecommand{\bibinfo}[2]{#2}
\providecommand{\eprint}[2][]{\url{#2}}

\bibitem[{\citenamefont{Leggett et~al.}(1987)\citenamefont{Leggett,
  Chakravarty, Dorsey, Fisher, Garg, and Zwerger}}]{Leggett:1987wk}
\bibinfo{author}{\bibfnamefont{A.~J.} \bibnamefont{Leggett}},
  \bibinfo{author}{\bibfnamefont{S.}~\bibnamefont{Chakravarty}},
  \bibinfo{author}{\bibfnamefont{A.~T.} \bibnamefont{Dorsey}},
  \bibinfo{author}{\bibfnamefont{M.~P.~A.} \bibnamefont{Fisher}},
  \bibinfo{author}{\bibfnamefont{A.}~\bibnamefont{Garg}}, \bibnamefont{and}
  \bibinfo{author}{\bibfnamefont{W.}~\bibnamefont{Zwerger}},
  \bibinfo{journal}{Rev. of Mod. Phys.} \textbf{\bibinfo{volume}{59}},
  \bibinfo{pages}{1} (\bibinfo{year}{1987}).

\bibitem[{\citenamefont{Breuer and Petruccione}(2002)}]{breuer:book}
\bibinfo{author}{\bibfnamefont{H.-P.} \bibnamefont{Breuer}} \bibnamefont{and}
  \bibinfo{author}{\bibfnamefont{F.}~\bibnamefont{Petruccione}},
  \emph{\bibinfo{title}{The Theory of Open Quantum Systems}}
  (\bibinfo{publisher}{Oxford Press}, \bibinfo{address}{Oxford},
  \bibinfo{year}{2002}).

\bibitem[{\citenamefont{U}(2012)}]{weiss:book}
\bibinfo{author}{\bibfnamefont{W.}~\bibnamefont{U}},
  \emph{\bibinfo{title}{Quantum Dissipative Systems}}
  (\bibinfo{publisher}{World Scientific}, \bibinfo{address}{Singapore},
  \bibinfo{year}{2012}).

\bibitem[{\citenamefont{Kryvohuz and Cao}(2005)}]{Kryvohuz:2005jf}
\bibinfo{author}{\bibfnamefont{M.}~\bibnamefont{Kryvohuz}} \bibnamefont{and}
  \bibinfo{author}{\bibfnamefont{J.~S.} \bibnamefont{Cao}},
  \bibinfo{journal}{Phys. Rev. Lett.} \textbf{\bibinfo{volume}{95}},
  \bibinfo{pages}{180405} (\bibinfo{year}{2005}).

\bibitem[{\citenamefont{Wu and Cao}(2001)}]{Wu:2001dd}
\bibinfo{author}{\bibfnamefont{J.}~\bibnamefont{Wu}} \bibnamefont{and}
  \bibinfo{author}{\bibfnamefont{J.~S.} \bibnamefont{Cao}},
  \bibinfo{journal}{J. Chem. Phys.} \textbf{\bibinfo{volume}{115}},
  \bibinfo{pages}{5381} (\bibinfo{year}{2001}).

\bibitem[{\citenamefont{Hsieh et~al.}(2012)\citenamefont{Hsieh, Shim,
  Korkusinski, and Hawrylak}}]{hsieh_rpp2012}
\bibinfo{author}{\bibfnamefont{C.-Y.} \bibnamefont{Hsieh}},
  \bibinfo{author}{\bibfnamefont{Y.-P.} \bibnamefont{Shim}},
  \bibinfo{author}{\bibfnamefont{M.}~\bibnamefont{Korkusinski}},
  \bibnamefont{and} \bibinfo{author}{\bibfnamefont{P.}~\bibnamefont{Hawrylak}},
  \bibinfo{journal}{Rep.Prog.Phys.} \textbf{\bibinfo{volume}{75}},
  \bibinfo{pages}{114501} (\bibinfo{year}{2012}).

\bibitem[{\citenamefont{Kloeffel and Loss}(2013)}]{Kloeffel:2013eg}
\bibinfo{author}{\bibfnamefont{C.}~\bibnamefont{Kloeffel}} \bibnamefont{and}
  \bibinfo{author}{\bibfnamefont{D.}~\bibnamefont{Loss}},
  \bibinfo{journal}{Anuu. Rev. Conden. Ma. P.} \textbf{\bibinfo{volume}{4}},
  \bibinfo{pages}{51} (\bibinfo{year}{2013}).

\bibitem[{\citenamefont{Gao et~al.}(2015)\citenamefont{Gao, Imamoglu, Bernien,
  and Hanson}}]{gao_natphoton2015}
\bibinfo{author}{\bibfnamefont{W.~B.} \bibnamefont{Gao}},
  \bibinfo{author}{\bibfnamefont{A.}~\bibnamefont{Imamoglu}},
  \bibinfo{author}{\bibfnamefont{H.}~\bibnamefont{Bernien}}, \bibnamefont{and}
  \bibinfo{author}{\bibfnamefont{R.}~\bibnamefont{Hanson}},
  \bibinfo{journal}{Nat. Photon.} \textbf{\bibinfo{volume}{9}},
  \bibinfo{pages}{363} (\bibinfo{year}{2015}).

\bibitem[{\citenamefont{Prokofiev and {Stamp, P. C.
  E.}}(2000)}]{Prokofiev:420342}
\bibinfo{author}{\bibfnamefont{N.~V.} \bibnamefont{Prokofiev}}
  \bibnamefont{and} \bibinfo{author}{\bibnamefont{{Stamp, P. C. E.}}},
  \bibinfo{journal}{Rep. Prog. Phys.} \textbf{\bibinfo{volume}{63}},
  \bibinfo{pages}{669} (\bibinfo{year}{2000}).

\bibitem[{\citenamefont{Hsieh and Cao}(2016)}]{hsieh_cao_2016}
\bibinfo{author}{\bibfnamefont{C.~Y.} \bibnamefont{Hsieh}} \bibnamefont{and}
  \bibinfo{author}{\bibfnamefont{J.S.}~\bibnamefont{Cao}} \bibinfo{journal}{Paper I.}
  (\bibinfo{year}{2016}).

\bibitem[{\citenamefont{Tang et~al.}(2016)\citenamefont{Tang, Ouyang, Gong,
  Wang, and Wu}}]{Tang:2016gh}
\bibinfo{author}{\bibfnamefont{Z.}~\bibnamefont{Tang}},
  \bibinfo{author}{\bibfnamefont{X.}~\bibnamefont{Ouyang}},
  \bibinfo{author}{\bibfnamefont{Z.}~\bibnamefont{Gong}},
  \bibinfo{author}{\bibfnamefont{H.}~\bibnamefont{Wang}}, \bibnamefont{and}
  \bibinfo{author}{\bibfnamefont{J.}~\bibnamefont{Wu}}, \bibinfo{journal}{J.
  Chem. Phys.} pp. \bibinfo{pages}{1--12} (\bibinfo{year}{2016}).

\bibitem[{\citenamefont{Stockburger and Grabert}(2002)}]{Stockburger:2002em}
\bibinfo{author}{\bibfnamefont{J.}~\bibnamefont{Stockburger}} \bibnamefont{and}
  \bibinfo{author}{\bibfnamefont{H.}~\bibnamefont{Grabert}},
  \bibinfo{journal}{Phys. Rev. Lett.} \textbf{\bibinfo{volume}{88}},
  \bibinfo{pages}{170407} (\bibinfo{year}{2002}).

\bibitem[{\citenamefont{Shao}(2004)}]{Shao:2004et}
\bibinfo{author}{\bibfnamefont{J.}~\bibnamefont{Shao}}, \bibinfo{journal}{J.
  Chem. Phys.} \textbf{\bibinfo{volume}{120}}, \bibinfo{pages}{5053}
  (\bibinfo{year}{2004}).

\bibitem[{\citenamefont{Lacroix}(2005)}]{Lacroix:2005in}
\bibinfo{author}{\bibfnamefont{D.}~\bibnamefont{Lacroix}},
  \bibinfo{journal}{Phys. Rev. A} \textbf{\bibinfo{volume}{72}},
  \bibinfo{pages}{013805} (\bibinfo{year}{2005}).

\bibitem[{\citenamefont{Tanimura}(2006)}]{Tanimura:2006ga}
\bibinfo{author}{\bibfnamefont{Y.}~\bibnamefont{Tanimura}},
  \bibinfo{journal}{J. Phys. Soc. Jpn.} \textbf{\bibinfo{volume}{75}},
  \bibinfo{pages}{082001} (\bibinfo{year}{2006}).

\bibitem[{\citenamefont{Makri}(1999)}]{makri99}
\bibinfo{author}{\bibfnamefont{N.}~\bibnamefont{Makri}}, \bibinfo{journal}{J.
  Phys. Chem. B} \textbf{\bibinfo{volume}{103}}, \bibinfo{pages}{2823}
  (\bibinfo{year}{1999}).

\bibitem[{\citenamefont{Moix and Cao}(2013)}]{Moix:2013jb}
\bibinfo{author}{\bibfnamefont{J.~M.} \bibnamefont{Moix}} \bibnamefont{and}
  \bibinfo{author}{\bibfnamefont{J.~S.} \bibnamefont{Cao}},
  \bibinfo{journal}{J. Chem. Phys.} \textbf{\bibinfo{volume}{139}},
  \bibinfo{pages}{134106} (\bibinfo{year}{2013}).

\bibitem[{\citenamefont{Cerrillo and Cao}(2014)}]{Cerrillo:2014gl}
\bibinfo{author}{\bibfnamefont{J.}~\bibnamefont{Cerrillo}} \bibnamefont{and}
  \bibinfo{author}{\bibfnamefont{J.~S.} \bibnamefont{Cao}},
  \bibinfo{journal}{Phys. Rev. Letts} \textbf{\bibinfo{volume}{112}},
  \bibinfo{pages}{110401} (\bibinfo{year}{2014}).

\bibitem[{\citenamefont{Jang et~al.}(2002)\citenamefont{Jang, Cao, and
  Silbey}}]{Jang:2002ds}
\bibinfo{author}{\bibfnamefont{S.}~\bibnamefont{Jang}},
  \bibinfo{author}{\bibfnamefont{J.~S.} \bibnamefont{Cao}}, \bibnamefont{and}
  \bibinfo{author}{\bibfnamefont{R.~J.} \bibnamefont{Silbey}},
  \bibinfo{journal}{J. Chem. Phys.} \textbf{\bibinfo{volume}{116}},
  \bibinfo{pages}{2705} (\bibinfo{year}{2002}).

\bibitem[{\citenamefont{Su{\'a}rez and Silbey}(1991)}]{Suarez:1991be}
\bibinfo{author}{\bibfnamefont{A.}~\bibnamefont{Su{\'a}rez}} \bibnamefont{and}
  \bibinfo{author}{\bibfnamefont{R.}~\bibnamefont{Silbey}},
  \bibinfo{journal}{J. Chem. Phys.} \textbf{\bibinfo{volume}{95}},
  \bibinfo{pages}{9115} (\bibinfo{year}{1991}).

\bibitem[{\citenamefont{Caldeira et~al.}(1993)\citenamefont{Caldeira, Neto, and
  de~Carvalho}}]{Caldeira:1993ud}
\bibinfo{author}{\bibfnamefont{A.~O.} \bibnamefont{Caldeira}},
  \bibinfo{author}{\bibfnamefont{A.~C.} \bibnamefont{Neto}}, \bibnamefont{and}
  \bibinfo{author}{\bibfnamefont{T.~O.} \bibnamefont{de~Carvalho}},
  \bibinfo{journal}{Phys. Rev. B} \textbf{\bibinfo{volume}{48}},
  \bibinfo{pages}{13974} (\bibinfo{year}{1993}).

\bibitem[{\citenamefont{Segal}(2014)}]{Segal:2014ws}
\bibinfo{author}{\bibfnamefont{D.}~\bibnamefont{Segal}}, \bibinfo{journal}{J.
  Chem. Phys.} \textbf{\bibinfo{volume}{140}}, \bibinfo{pages}{164110}
  (\bibinfo{year}{2014}).

\bibitem[{\citenamefont{Wang and Shao}(2012)}]{Wang:2012kk}
\bibinfo{author}{\bibfnamefont{H.}~\bibnamefont{Wang}} \bibnamefont{and}
  \bibinfo{author}{\bibfnamefont{J.}~\bibnamefont{Shao}}, \bibinfo{journal}{J.
  Chem. Phys.} \textbf{\bibinfo{volume}{137}}, \bibinfo{pages}{22A504}
  (\bibinfo{year}{2012}).

\bibitem[{\citenamefont{Gelman et~al.}(2004)\citenamefont{Gelman, Koch, and
  Kosloff}}]{Gelman:2004ji}
\bibinfo{author}{\bibfnamefont{D.}~\bibnamefont{Gelman}},
  \bibinfo{author}{\bibfnamefont{C.~P.} \bibnamefont{Koch}}, \bibnamefont{and}
  \bibinfo{author}{\bibfnamefont{R.}~\bibnamefont{Kosloff}},
  \bibinfo{journal}{J. Chem. Phys.} \textbf{\bibinfo{volume}{121}},
  \bibinfo{pages}{661} (\bibinfo{year}{2004}).

\bibitem[{\citenamefont{Zhang et~al.}(2015)\citenamefont{Zhang, Xu, Zheng, and
  Yan}}]{Zhang:2015jl}
\bibinfo{author}{\bibfnamefont{H.-D.} \bibnamefont{Zhang}},
  \bibinfo{author}{\bibfnamefont{R.-X.} \bibnamefont{Xu}},
  \bibinfo{author}{\bibfnamefont{X.}~\bibnamefont{Zheng}}, \bibnamefont{and}
  \bibinfo{author}{\bibfnamefont{Y.}~\bibnamefont{Yan}}, \bibinfo{journal}{J.
  Chem. Phys.} \textbf{\bibinfo{volume}{142}}, \bibinfo{pages}{024112}
  (\bibinfo{year}{2015}).

\bibitem[{\citenamefont{L{\"u} and Zheng}(2009)}]{Lu:2009hd}
\bibinfo{author}{\bibfnamefont{Z.}~\bibnamefont{L{\"u}}} \bibnamefont{and}
  \bibinfo{author}{\bibfnamefont{H.}~\bibnamefont{Zheng}}, \bibinfo{journal}{J.
  Chem. Phys.} \textbf{\bibinfo{volume}{131}}, \bibinfo{pages}{134503}
  (\bibinfo{year}{2009}).

\bibitem[{\citenamefont{Bortz and Stolze}(2007)}]{Bortz:2007ku}
\bibinfo{author}{\bibfnamefont{M.}~\bibnamefont{Bortz}} \bibnamefont{and}
  \bibinfo{author}{\bibfnamefont{J.}~\bibnamefont{Stolze}},
  \bibinfo{journal}{Phys. Rev. B} \textbf{\bibinfo{volume}{76}},
  \bibinfo{pages}{014304} (\bibinfo{year}{2007}).

\bibitem[{\citenamefont{Wang et~al.}(2013)\citenamefont{Wang, Guo, and
  Zhou}}]{Wang:2013jx}
\bibinfo{author}{\bibfnamefont{Z.~H.} \bibnamefont{Wang}},
  \bibinfo{author}{\bibfnamefont{Y.}~\bibnamefont{Guo}}, \bibnamefont{and}
  \bibinfo{author}{\bibfnamefont{D.~L.} \bibnamefont{Zhou}},
  \bibinfo{journal}{Eur. Phys. J. D} \textbf{\bibinfo{volume}{67}},
  \bibinfo{pages}{218} (\bibinfo{year}{2013}).

\bibitem[{\citenamefont{Chen et~al.}(2007)\citenamefont{Chen, Bergman, and
  Balents}}]{Chen:2007kra}
\bibinfo{author}{\bibfnamefont{G.}~\bibnamefont{Chen}},
  \bibinfo{author}{\bibfnamefont{D.~L.} \bibnamefont{Bergman}},
  \bibnamefont{and} \bibinfo{author}{\bibfnamefont{L.}~\bibnamefont{Balents}},
  \bibinfo{journal}{Physical Review B} \textbf{\bibinfo{volume}{76}},
  \bibinfo{pages}{045312} (\bibinfo{year}{2007}).

\bibitem[{\citenamefont{Seifert et~al.}(2016)\citenamefont{Seifert, Bleicker,
  Schering, Faribault, and Uhrig}}]{Seifert:2016vx}
\bibinfo{author}{\bibfnamefont{U.}~\bibnamefont{Seifert}},
  \bibinfo{author}{\bibfnamefont{P.}~\bibnamefont{Bleicker}},
  \bibinfo{author}{\bibfnamefont{P.}~\bibnamefont{Schering}},
  \bibinfo{author}{\bibfnamefont{A.}~\bibnamefont{Faribault}},
  \bibnamefont{and} \bibinfo{author}{\bibfnamefont{G.~S.} \bibnamefont{Uhrig}},
  \bibinfo{journal}{Phys. Rev. B} \textbf{\bibinfo{volume}{94}},
  \bibinfo{pages}{094308} (\bibinfo{year}{2016}).

\bibitem[{\citenamefont{Breuer et~al.}(2004)\citenamefont{Breuer, Burgarth, and
  Petruccione}}]{Breuer:2004vi}
\bibinfo{author}{\bibfnamefont{H.-P.} \bibnamefont{Breuer}},
  \bibinfo{author}{\bibfnamefont{D.}~\bibnamefont{Burgarth}}, \bibnamefont{and}
  \bibinfo{author}{\bibfnamefont{F.}~\bibnamefont{Petruccione}},
  \bibinfo{journal}{Phys. Rev. B} \textbf{\bibinfo{volume}{70}},
  \bibinfo{pages}{045323. 11 p} (\bibinfo{year}{2004}).

\bibitem[{\citenamefont{L{\'o}pez-L{\'o}pez
  et~al.}(2011)\citenamefont{L{\'o}pez-L{\'o}pez, Martinazzo, and
  Nest}}]{LopezLopez:2011en}
\bibinfo{author}{\bibfnamefont{S.}~\bibnamefont{L{\'o}pez-L{\'o}pez}},
  \bibinfo{author}{\bibfnamefont{R.}~\bibnamefont{Martinazzo}},
  \bibnamefont{and} \bibinfo{author}{\bibfnamefont{M.}~\bibnamefont{Nest}},
  \bibinfo{journal}{J. Chem. Phys.} \textbf{\bibinfo{volume}{134}},
  \bibinfo{pages}{094102} (\bibinfo{year}{2011}).

\bibitem[{\citenamefont{Lloyd}(1996)}]{Lloyd_sci96}
\bibinfo{author}{\bibfnamefont{S.}~\bibnamefont{Lloyd}},
  \bibinfo{journal}{Science} \textbf{\bibinfo{volume}{273}},
  \bibinfo{pages}{1073} (\bibinfo{year}{1996}).

\bibitem[{\citenamefont{Krovi et~al.}(2007)\citenamefont{Krovi, Oreshkov,
  Ryazanov, and Lidar}}]{Krovi:2007iu}
\bibinfo{author}{\bibfnamefont{H.}~\bibnamefont{Krovi}},
  \bibinfo{author}{\bibfnamefont{O.}~\bibnamefont{Oreshkov}},
  \bibinfo{author}{\bibfnamefont{M.}~\bibnamefont{Ryazanov}}, \bibnamefont{and}
  \bibinfo{author}{\bibfnamefont{D.}~\bibnamefont{Lidar}},
  \bibinfo{journal}{Phys. Rev. A} \textbf{\bibinfo{volume}{76}},
  \bibinfo{pages}{052117} (\bibinfo{year}{2007}).

\bibitem[{\citenamefont{Torrontegui and Kosloff}(2016)}]{torron_kosloff_njp16}
\bibinfo{author}{\bibfnamefont{E.}~\bibnamefont{Torrontegui}} \bibnamefont{and}
  \bibinfo{author}{\bibfnamefont{R.}~\bibnamefont{Kosloff}},
  \bibinfo{journal}{New J. Phys.} \textbf{\bibinfo{volume}{18}},
  \bibinfo{pages}{093001} (\bibinfo{year}{2016}).

\bibitem[{\citenamefont{Dobrovitski and De~Raedt}(2003)}]{Dobrovitski:602306}
\bibinfo{author}{\bibfnamefont{V.~V.} \bibnamefont{Dobrovitski}}
  \bibnamefont{and} \bibinfo{author}{\bibfnamefont{H.~A.}
  \bibnamefont{De~Raedt}}, \bibinfo{journal}{Phys. Rev. E}
  \textbf{\bibinfo{volume}{67}}, \bibinfo{pages}{056702. 8 p}
  (\bibinfo{year}{2003}).

\bibitem[{\citenamefont{Al-Hassanieh et~al.}(2006)\citenamefont{Al-Hassanieh,
  Dobrovitski, Dagotto, and Harmon}}]{AlHassanieh:2006kr}
\bibinfo{author}{\bibfnamefont{K.}~\bibnamefont{Al-Hassanieh}},
  \bibinfo{author}{\bibfnamefont{V.~V.} \bibnamefont{Dobrovitski}},
  \bibinfo{author}{\bibfnamefont{E.}~\bibnamefont{Dagotto}}, \bibnamefont{and}
  \bibinfo{author}{\bibfnamefont{B.}~\bibnamefont{Harmon}},
  \bibinfo{journal}{Phys. Rev. Lett.} \textbf{\bibinfo{volume}{97}},
  \bibinfo{pages}{037204} (\bibinfo{year}{2006}).

\bibitem[{\citenamefont{Stanek et~al.}(2013)\citenamefont{Stanek, Raas, and
  Uhrig}}]{stanek:2013}
\bibinfo{author}{\bibfnamefont{D.}~\bibnamefont{Stanek}},
  \bibinfo{author}{\bibfnamefont{C.}~\bibnamefont{Raas}}, \bibnamefont{and}
  \bibinfo{author}{\bibfnamefont{G.~S.} \bibnamefont{Uhrig}},
  \bibinfo{journal}{Phys. Rev. B} \textbf{\bibinfo{volume}{88}},
  \bibinfo{pages}{155305} (\bibinfo{year}{2013}).

\bibitem[{\citenamefont{Shenvi et~al.}(2005)\citenamefont{Shenvi, de~Sousa, and
  Whaley}}]{shenvi_desousa_prb2005}
\bibinfo{author}{\bibfnamefont{N.}~\bibnamefont{Shenvi}},
  \bibinfo{author}{\bibfnamefont{R.}~\bibnamefont{de~Sousa}}, \bibnamefont{and}
  \bibinfo{author}{\bibfnamefont{K.~B.} \bibnamefont{Whaley}},
  \bibinfo{journal}{Phys. Rev. B} \textbf{\bibinfo{volume}{71}},
  \bibinfo{pages}{224411} (\bibinfo{year}{2005}).

\bibitem[{\citenamefont{Schliemann et~al.}(2003)\citenamefont{Schliemann,
  Khaetskii, and Loss}}]{loss_2003}
\bibinfo{author}{\bibfnamefont{J.}~\bibnamefont{Schliemann}},
  \bibinfo{author}{\bibfnamefont{A.}~\bibnamefont{Khaetskii}},
  \bibnamefont{and} \bibinfo{author}{\bibfnamefont{D.}~\bibnamefont{Loss}},
  \bibinfo{journal}{J. Phys. C} \textbf{\bibinfo{volume}{15}},
  \bibinfo{pages}{R1809} (\bibinfo{year}{2003}).

\bibitem[{\citenamefont{Zhang et~al.}(2006)\citenamefont{Zhang, Dobrovitski,
  Al-Hassanieh, Dagotto, and Harmon}}]{zhang_dobrovitski_prb2006}
\bibinfo{author}{\bibfnamefont{W.}~\bibnamefont{Zhang}},
  \bibinfo{author}{\bibfnamefont{V.~V.} \bibnamefont{Dobrovitski}},
  \bibinfo{author}{\bibfnamefont{K.}~\bibnamefont{Al-Hassanieh}},
  \bibinfo{author}{\bibfnamefont{E.}~\bibnamefont{Dagotto}}, \bibnamefont{and}
  \bibinfo{author}{\bibfnamefont{B.}~\bibnamefont{Harmon}},
  \bibinfo{journal}{Phys. Rev. B} \textbf{\bibinfo{volume}{74}},
  \bibinfo{pages}{205313} (\bibinfo{year}{2006}).

\bibitem[{\citenamefont{Bhaktavatsala~Rao and Kurizki}(2011)}]{rao-pra2011}
\bibinfo{author}{\bibfnamefont{D.~D.} \bibnamefont{Bhaktavatsala~Rao}}
  \bibnamefont{and} \bibinfo{author}{\bibfnamefont{G.}~\bibnamefont{Kurizki}},
  \bibinfo{journal}{Phys. Rev. A} \textbf{\bibinfo{volume}{83}},
  \bibinfo{pages}{032105} (\bibinfo{year}{2011}).

\bibitem[{\citenamefont{Pach{\'o}n and Brumer}(2011)}]{Pachon:2011ea}
\bibinfo{author}{\bibfnamefont{L.~A.} \bibnamefont{Pach{\'o}n}}
  \bibnamefont{and} \bibinfo{author}{\bibfnamefont{P.}~\bibnamefont{Brumer}},
  \bibinfo{journal}{J. Phys. Chem. Lett.} \textbf{\bibinfo{volume}{2}},
  \bibinfo{pages}{2728} (\bibinfo{year}{2011}).

\bibitem[{\citenamefont{Wang et~al.}(2015)\citenamefont{Wang, Ren, and
  Cao}}]{Wang:2015jz}
\bibinfo{author}{\bibfnamefont{C.}~\bibnamefont{Wang}},
  \bibinfo{author}{\bibfnamefont{J.}~\bibnamefont{Ren}}, \bibnamefont{and}
  \bibinfo{author}{\bibfnamefont{J.~S.} \bibnamefont{Cao}},
  \bibinfo{journal}{Sci. Rep.} p. \bibinfo{pages}{1187} (\bibinfo{year}{2015}).

\end{thebibliography}

\end{document}